\documentclass{jfm}
\usepackage{graphicx}
\usepackage{newtxtext}
\usepackage{newtxmath}
\usepackage{color}
\usepackage{natbib}
\usepackage{hyperref}
\hypersetup{
    colorlinks = true,
    urlcolor   = blue,
    citecolor  = black,
}
\usepackage{amssymb}

\newcommand{\RomanNumeralCaps}[1]
\linenumbers

\renewcommand{\vec}{\boldsymbol}

\newcommand\NU{\mbox{\textit{Nu}}}
\newcommand\RA{\mbox{\textit{Ra}}}
\newcommand{\PR}{\mbox{\textit{Pr}}}

\def\Gamma  {\varGamma}

\title{Oscillatory large-scale circulation in liquid-metal thermal convection and its structural unit}

\author{Andrei Teimurazov\aff{1}
 \corresp{andrei.teimurazov@ds.mpg.de},
 Sanjay Singh\aff{2}
 \corresp{s.singh@hzdr.de},
 Sylvie Su\aff{2}
 \corresp{s.su@hzdr.de},
 Sven Eckert\aff{2}
 \corresp{s.eckert@hzdr.de},
 Olga Shishkina\aff{1}
 \corresp{olga.shishkina@ds.mpg.de},
\and Tobias Vogt\aff{2}
 \corresp{t.vogt@hzdr.de}
}

\affiliation{\aff{1}Max Planck Institute for Dynamics and Self-Organization, Am Fassberg 17, 37077 G\"ottingen, Germany
\aff{2}Institute of Fluid Dynamics, Helmholtz-Zentrum Dresden-Rossendorf, 01328 Dresden, Germany}

\shorttitle{The structural unit of oscillatory large-scale circulation}
\shortauthor{A. Teimurazov, S. Singh, S. Su,  S. Eckert, O. Shishkina and T. Vogt }

\begin{document}
\maketitle

\begin{abstract}
In Rayleigh--B\'enard convection (RBC), the size of a flow domain and its aspect ratio $\Gamma$ (a ratio between the spatial length and height of the domain) affect the shape of the large-scale circulation (LSC). For some aspect ratios, the flow dynamics include a three-dimensional oscillatory mode known as a jump-rope vortex (JRV), however, the effects of varying aspect ratios on this mode are not well investigated. In this paper, we study these aspect-ratio effects in liquid metals, for a low Prandtl number $\Pran=0.03$. Direct numerical simulations and experiments are carried out for a Rayleigh number range $2.9 \times 10^4 \leq \RA \leq 1.6 \times 10^6$ and square cuboid domains with $\Gamma=2$, $2.5$, $3$ and $5$.
Our study demonstrates that a repeating pattern of a JRV encountered at an aspect ratio $\Gamma\approx2.5$ is the basic structural unit that builds up to a lattice of interlaced JRVs at the largest aspect ratio. The size of the domain determines how many structural units are self-organized within the domain; the number of the realized units is expected to scale as $\Gamma^2$ with sufficiently large and growing $\Gamma$. We find the oscillatory modes for all investigated $\Gamma$, however, they are more pronounced for $\Gamma=2.5$ and $\Gamma=5$. Future studies for large-aspect ratio domains of different shapes would enhance our understanding of how the JRVs adjust and reorganize at such scaled-up geometries, and answer the question of whether they are indeed the smallest superstructure units. 
\end{abstract}

\section{Introduction}
\label{sec:intro}
Thermal convection manifests not only in various geo- and astrophysical systems, but also in smaller-scale phenomena ranging from industrial processes to our daily lives such as household heating. 
Given its importance, natural thermal convection has been the subject of intensive research for over a century \citep{Benard1900, Rayleigh1916}. 
Investigation of thermal convection in low-Prandtl-number fluids (Prandtl numbers $\Pran \ll 1$) is of particular importance for a better understanding of convection on surfaces of stars, where $\Pran$ can be as low as $10^{-8}$ \citep{Spiegel1962, Hanasoge2016},
and, in case of liquid metals, for numerous technical applications, e.g. the advancement of cooling technology
\citep[see, e.g.,][]{Scheel2013, Frick2015, Schumacher2015, Scheel2016, Heinzel2017, Teimurazov2017a, Zuerner2019, Pandey2022, Zwirner2022}.

Natural thermal convection occurs in a fluid layer due to a temperature difference imposed at its surfaces.
Here, the orientation of the fluid layer surfaces with respect to the gravity vector plays an important role
\citep[see, e.g., ][]{Shishkina2016b, Teimurazov2017, Zuerner2019, Zwirner2020a, Teimurazov2021}.
One of the classical and probably the most intensively investigated configurations of natural thermal convection is Rayleigh--B\'{e}nard convection (RBC) \citep{Benard1900, Rayleigh1916, Bodenschatz2000, Ahlers2009, Chilla2012}.
In RBC, the heated and cooled surfaces are placed orthogonal to the gravity vector, and the fluid layer is heated from below and cooled from above.
Thermal expansion causes warm fluid to rise and cool fluid to sink. 
At sufficiently large Rayleigh number, $\RA\equiv~\alpha g \Delta H^3 / (\kappa \nu)$, the resulting turbulent convective flow self-organises through an inverse energy cascade from small-scale thermal turbulence to large flow structures. 
Here, $\alpha$ is the isobaric thermal expansion coefficient, $\nu$ is the kinematic viscosity, $\kappa$ is the thermal diffusivity, $\Delta$ is the temperature difference between the heated and cooled surfaces, $H$ is the distance between these surfaces  (i.e., the height of the container) and $g$ denotes the acceleration due to gravity.
The energy of small scales is directly transferred to large scales via three-dimensional modes, and is different from the classical two-dimensional inverse energy cascade \citep{Ecke2023, Boffetta2012}.

At sufficiently large $\RA$, the flow is self-organised into a large-scale circulation (LSC), or a turbulent thermal wind, the concept of which is an important ingredient in the heat and momentum transport theory \citep{Grossmann2000, Grossmann2001, Grossmann2011}, and boundary-layer theory for natural thermal convection \citep{Shishkina2015, Ching2019,  Tai2021}.  
The resulting flow structures strongly depend on the Rayleigh number $\RA$, which is a measure of the thermal forcing that drives convection in the system, and on the Prandtl number $\Pran\equiv\nu/\kappa$, which describes the diffusive properties of the considered fluid \citep{Ahlers2009}. 
In addition, the geometric characteristics of the container, especially the shape of the container and, in particular, the aspect ratio 
 $\Gamma$ of its spatial length $L$ and height $H$, $\Gamma\equiv~L/H$, influence the global flow structure and the mean characteristics of the flow \citep{Shishkina2021, Ahlers2022}.

Turbulent RBC in a cylindrical container with equal height and diameter (aspect ratio $\Gamma=1$) is the most extensively studied. For containers with $\Gamma\approx1$, the principle structure of the LSC can be delineated as follows.
There exists a vertical plane (called the LSC-plane), in which the LSC is observed as a big domain-filling roll with two smaller secondary rolls in the corners \citep{Sun2005b, Ahlers2009, Chilla2012}, while in the orthogonal vertical plane, the LSC for this geometry of the container is seen as a four-roll structure, with an inflow at mid-height \citep{Shishkina2014}.
Not only the LSC is generally unsteady, but also the LSC path can exhibit dynamic behaviour. 
Thus in containers with $\Gamma\approx1$, the LSC can display various modes of periodic or chaotic oscillations which can take the form of sloshing, precession, and torsion \citep{Cioni1997, Xi2004, Funfschilling2004, Sun2005, Xi2006, Brown2006, Brown2007, Xi2007, Xi2008, Funfschilling2008, Zhou2009, Brown2009, Sugiyama2010, Assaf2011, Stevens2011, Wagner2012, Sakievich2016, Sakievich2020, Cheng2022}.
The sloshing mode is associated with the motion of the LSC-plane in the radial direction, while the precession and torsion modes are related to the azimuthal motion of the LSC-plane \citep{Cheng2022, Horn2022}.
In the precession mode, the entire LSC-plane drifts in the azimuthal direction, while in the torsion mode, the azimuthal motion of the LSC-plane in the upper half of the container is generally in the opposite direction to the motion of the LSC-plane in the lower half of the container. 

In slender containers with the aspect ratio $\Gamma<1$, a single big-roll structure of the LSC is not as stable as in the case of $\Gamma=1$ \citep{Xi2008, Weiss2011a, Weiss2013, Zwirner2020, Schindler2022}.
For $\Gamma<1$, the turbulent LSC can be formed of several dynamically changing convective rolls that are stacked on top of each other \citep{Poel2011, Poel2012, Zwirner2020}.
The mechanism which causes the twisting and breaking of a single-roll LSC into multiple rolls is the
elliptical instability \citep{Zwirner2020}.
In the case of $\Gamma<1$, the heat and momentum transports, which are represented by the Nusselt number $Nu$ and Reynolds number $Re$, are always stronger for a smaller number of the rolls that form the LSC.
This was proven in experiments for $\Gamma=1/2$ \citep{Weiss2011a, Weiss2013, Xi2008}, and simulations for $\Gamma=1/5$ \citep{Zwirner2018, Zwirner2020}.

\begin{figure}
%figure 1
\centerline{\includegraphics[scale=1.0]{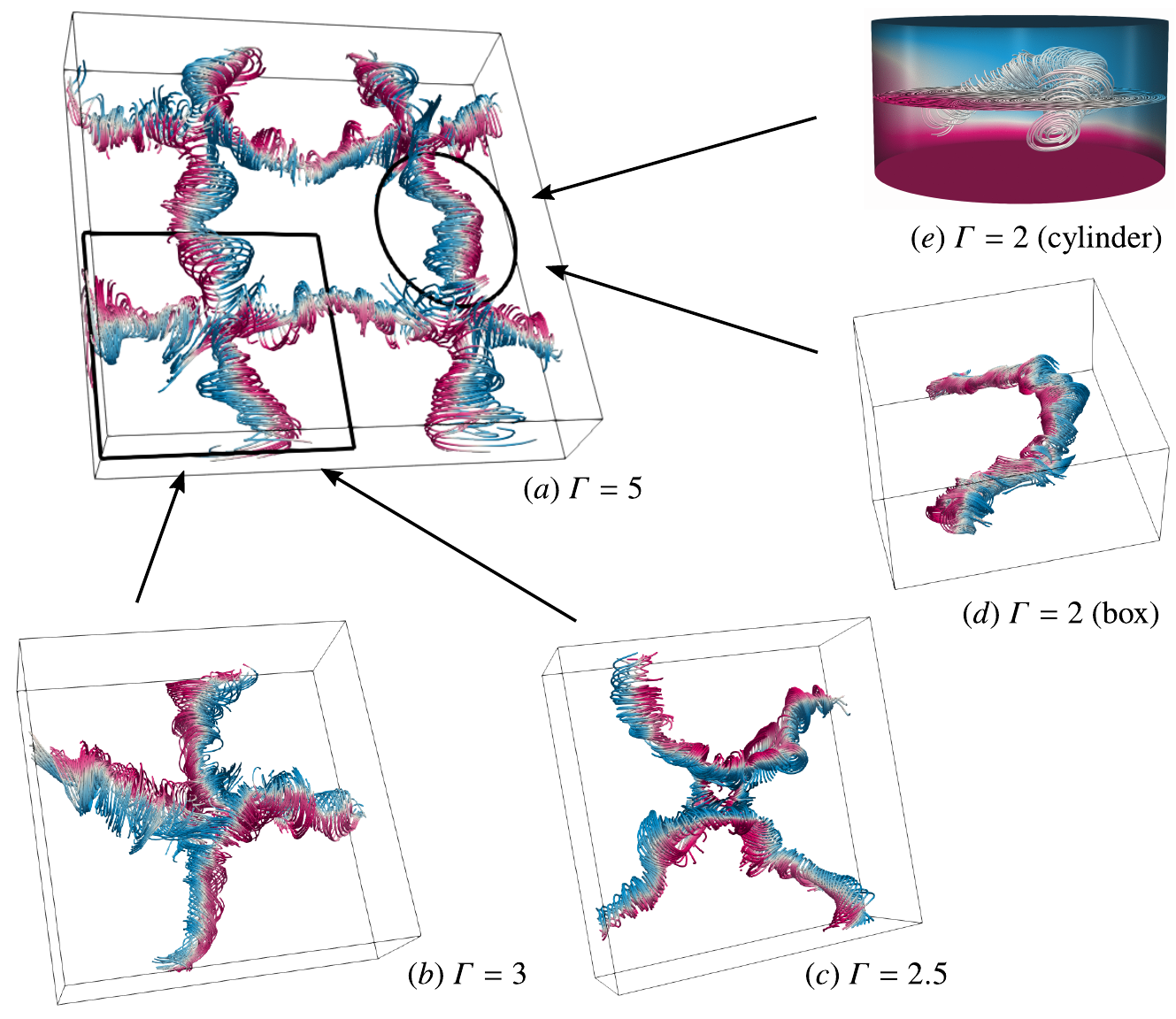}}
\caption{
Phase-averaged streamlines in Rayleigh--B\'enard convection for $\Pran=0.03$, $\RA=10^6$,  as obtained in direct numerical simulations for different aspect ratios (\textit{a}) $\Gamma=5$, (\textit{b}) $\Gamma=3$, (\textit{c}) $\Gamma=2.5$ and (\textit{d}) $\Gamma=2$ for square cuboid domains (new simulations) and (\textit{e}, adapted from \cite{Vogt2018}, available under a \href{https://creativecommons.org/licenses/by-nc-nd/4.0/}{CC BY-NC-ND 4.0}) a cylindrical domain with $\Gamma=2$.
These streamlines envelope the oscillating vortex, and the colour scale is according to the vertical velocity component $u_z$.
Blue (red) colour corresponds to a negative (positive) value of $u_z$, indicating a downward (upward) direction of the flow. 
The structures in the lower aspect ratio cases ($\Gamma=2$, 2.5 and 3) are building units of the structure formed within the largest aspect ratio case ($\Gamma=5$).
}
\label{JRV_family}
\end{figure}

By contrast, for wide containers with $\Gamma > 1$, the more rolls of the LSC mean the more efficient heat transport \citep{Poel2011, Poel2012, Wang2020}, also in highly turbulent cases. 
For $\Gamma > 1$, the rolls or roll-like structures are attached to each other and aligned in horizontal directions \citep{Hartlep2003, Hardenberg2008, Emran2015}.
In the two-dimensional case, the range of possible aspect ratios of particular convective rolls and, hence, the total number of the rolls in any confined domain, are restricted, and there exist quite accurate theoretical estimates for the lower and upper bounds of possible aspect ratios of the rolls \citep{Wang2020, Shishkina2021}. 

For three-dimensional domains, the typical length scales of the self-organised coherent turbulent flow structures are not yet well-studied and their accurate prediction remains an unsolved problem so far.
These flow structures can be identified as turbulent superstructures \citep{Stevens2018, Pandey2018, Green2020, Krug2020, Berghout2021}, since their lifetime is much larger than the free-fall time, and their length scales are generally larger than the typical length scale in RBC, which is the height of the container $H$.
Several studies suggest that the characteristic length scale of these coherent turbulent large-scale flow structures increases with growing $\RA$, see, e.g., \citet{Fitzjarrald1976, Hartlep2003, Pandey2018, Akashi2019, Krug2020}. 
Depending on the considered parametric space of the $\RA$-, $\Pran$- and $\Gamma$ in different studies, different preferable length scales of the turbulent superstructures are reported, which are always larger that the container height $H$.
Thus the values of order $10H$ \citep{Busse1994}, or between $6H$ and $7H$ \citep{Hartlep2003, Pandey2018, Stevens2018} were proposed.
Although the typical horizontal wavelengths of the turbulent superstructures generally grow with $\RA$, they tend to decrease with decreasing Prandtl number \citep{Pandey2018}. 
This fact is pretty remarkable, since decreasing $\Pran$ is usually associated with even stronger turbulence and therefore one might expect a certain similarity to the situation when $\RA$ is increased.

Recent laboratory and numerical experiments show that in an intermediate range of moderate aspect ratios, $\Gamma \gtrsim 1.4$, the LSC displays a low-frequency oscillatory dynamics \citep{Vogt2018, Horn2022, Akashi2022, Cheng2022}. 
The precession, torsional and sloshing dynamics of the LSC, which dominates at $\Gamma=1$, is replaced by a mode which can be described as a jump rope vortex (JRV). 
In this flow pattern, a curved vortex is formed, which swirls around the cell centre in the direction opposite to the LSC direction, resembling the motion of a swirling jump rope, see figure~\ref{JRV_family}e. 
This phenomenon was first demonstrated for liquid metal convection in a cylinder with aspect ratio $\Gamma=2$ \citep{Vogt2018}. 
Numerical simulations showed that the JRV exists also for a cylindrical container of the aspect ratio $\Gamma=\sqrt{2}$ and that the JRV structure is present not only in low-$\Pran$ liquid metal convection, but also in water at $\Pran = 4.8$. 
This has been confirmed in several other experiments and simulations of comparable aspect ratios for both water and liquid metal \citep{Horn2022, Cheng2022, Li2022}.  

Flow measurements in containers of different shapes such as a cuboid domain with $\Gamma = 5$ \citep{Akashi2022} showed that the strongly oscillating velocity and temperature fields could also be attributed to the presence of the JRV-like structures. 
However, instead of only one vortex, four JRVs interlaced in that case. 
The ends of the JRVs cross perpendicularly at a certain point in space (see figure~\ref{JRV_family}\textit{a}). 
Here, the detached (opposite) JRVs oscillate $\pi$ out of phase, whereas adjacent JRVs do so with a lag of $\pi/2$. 
\cite{Akashi2022} demonstrated that the JRVs can form a lattice structure of different vortices, which determines a fundamental flow mode that for the considered combinations of $\RA$ and $\Pran$ can dominate the dynamics at moderate, and possibly also at very large aspect ratios.
 
Although JRVs have also been detected in water with moderate $\Pran \approx 5$, the liquid metal offers a number of advantages for such studies. 
The velocity field in liquid metal convection is strongly inertia dominated due to its low viscosity and high density. 
As a result, the JRV induced oscillations reach much stronger amplitudes than in water or air. 
While the velocity field in low-$\Pran$ liquid metal at comparable temperature gradients is significantly more turbulent than that of water or air, the temperature field  exhibits considerably more coherence than the velocity field due to the large thermal diffusivity. 
Thus, the JRV-induced oscillations can be detected both in the velocity field and in the temperature field very well. 
As such, liquid metals are well suited to investigate the JRV-like flow dynamics.

The objective of the present work is to investigate in more detail the aspect ratio and geometry dependence of the three-dimensional oscillatory JRV-like large-scale circulation in liquid-metal thermal convection. 
In particular, how increasing aspect ratios result in a lattice of oscillatory flow pattern via the formation of JRVs, starting from the smallest structural building block to that of the more interlaced JRVs at a higher aspect ratio.
To this end, we study the LSC dynamics in RBC of liquid metal with $\Pran=0.03$ in square cuboids with different aspect ratios, which vary from 2 to 5, using both experimental and numerical approaches.

\section{Methods}
\label{sec:setup}

\subsection{Direct numerical simulations}

\begin{table}
	\begin{center}
		\def~{\hphantom{0}}
		\small
		\begin{tabular}{lcccccllcccc}
			$\Gamma$ & $\Pran$ &$\RA$ & $\ N_x\ $ & $\ N_y\ $ & $\ N_z\ $& $\mathcal{N}_\theta$&$\	\mathcal{N}_{\text{v}}\ $ & $ \delta_\theta/H$ & $\delta_{\text{v}}/H$ & $h_{\text{K}}$ & $h_{\text{DNS}}/h_{\text{K}}$\\ \hline
			2  & $0.03$ & $1.0 \times 10^6$ & 600  & 600  & 300 & 35 & 10 & $9.9\times10^{-2}$ & $2.5\times10^{-2}$ & $3.9\times10^{-3}$ & 0.94 \\ \hline
			2.5& $0.03$ & $1.2 \times 10^5$ & 750  & 750  & 300 & 58 & 16 & $1.7\times10^{-1}$ & $4.4\times10^{-2}$ & $7.9\times10^{-3}$ & 0.47 \\
			&        & $1.0 \times 10^6$ & 750  & 750  & 300 & 33 & 9 & $9.4\times10^{-2}$ & $2.4\times10^{-2}$ & $3.8\times10^{-3}$ & 0.97 \\ \hline
			3  & $0.03$ & $1.0 \times 10^5$ & 720  & 720  & 240 & 47 & 13 & $1.7\times10^{-1}$ & $4.4\times10^{-2}$ & $8.3\times10^{-3}$ & 0.56 \\
			&        & $4.05 \times 10^5$ & 780  & 780  & 260 & 46 & 15 & $1.2\times10^{-1}$ & $3.2\times10^{-2}$ & $5.2\times10^{-3}$ & 0.95 \\ 
			&        & $1.0 \times 10^6$ & 900  & 900  & 300 & 34 & 9 & $9.6\times10^{-2}$ & $2.4\times10^{-2}$ & $3.8\times10^{-3}$ & 0.96 \\   \hline
			5  & $0.03$ & $1.2 \times 10^5$ & 1500 & 1500 & 300 & 56 & 16 & $1.7\times10^{-1}$ & $4.2\times10^{-2}$ & $7.8\times10^{-3}$ & 0.47 \\  
			&        & $1.0 \times 10^6$ & 1500 & 1500 & 300 & 33 & 9 & $9.3\times10^{-2}$ & $2.4\times10^{-2}$ & $3.8\times10^{-3}$ & 0.97 \\
			
		\end{tabular}
		\caption{Details on the conducted DNS, including the number of nodes $N_x$, $N_y$, $N_z$ in the directions $x$, $y$ and $z$, respectively;
			the number of the nodes within the thermal boundary layer, $\mathcal{N}_{\theta}$, 
			and within the viscous boundary layer, $\mathcal{N}_{\text{v}}$, 
			the relative thickness of the thermal boundary layer, $\delta_{\theta}/H$, and the viscous boundary layer $\delta_{\text{v}}/H$; 
			the Kolmogorov microscale, $h_{\text{K}}$,
			and the relative mean grid stepping, $h_{\text{DNS}}/h_{\text{K}}$.
		}
		\label{TAB}
	\end{center}
\end{table}  

Thermal convection under the assumption of the Oberbeck--Boussinesq approximation is described by the following Navier--Stokes, energy, and continuity equations:
\begin{eqnarray}
D_t \vec{u} &=&  \nu\vec{\nabla}^2 \vec{u} - \vec{\nabla} p + \alpha g (T-T_0) \vec{e}_z, \label{eq:NS1}\\
D_t T &=&  \kappa \vec{\nabla}^2 T,\label{eq:NS2}\\
\vec{\nabla} \cdot \vec{u} &=& 0. \label{eq:NS3}
\end{eqnarray}
Here, $D_{t}$ denotes the substantial derivative, $\boldsymbol{u} = (u_x, u_y, u_z)$ is the velocity vector field, 
$p$ is the reduced kinematic pressure, 
$T$ the temperature, 
$T_0=(T_++T_-)/2$ is the arithmetic mean of the top ($T_-$) and bottom ($T_+$) temperatures,
$\vec{e}_z$ is the unit vector that points upward.
The considered domain is a square cuboid with the height $H$ and equal width $W$ and length $L$, $W=L$, 
so that the domain aspect ratio equals $\Gamma\equiv L/H$.
The system (\ref{eq:NS1})--(\ref{eq:NS3}) is closed by the following boundary conditions: no-slip for the velocity at all boundaries, $\vec{u}=0$, 
constant temperatures at the end-face of the box, i.e., 
$T=T_+$ at the bottom plate at $z=0$ and 
$T=T_-$ at the top plate at $z=H$,
and adiabatic boundary condition at the side walls, $\partial T/\partial \vec{n}=0$, where $\vec{n}$ is the vector orthogonal to the surface.
Equations (\ref{eq:NS1})--(\ref{eq:NS3}) are non-dimensionalised by using the height $H$,
the free-fall velocity $u_{f\!f}$, the free-fall time $t_{f\!f}$,
and the temperature difference between the heated plate and the cooled plate, $\Delta$, 
\begin{eqnarray}
    u_{f\!f}\equiv(\alpha gH\Delta)^{1/2},\qquad
    t_{f\!f}\equiv H/u_{f\!f},\qquad
    \Delta\equiv T_+-T_-,
\label{dimless}    
\end{eqnarray}
as units of length, velocity, time and temperature, respectively.

The resulting dimensionless equations are solved numerically using the latest version \citep{Reiter2022, Reiter2021a}
of the direct numerical solver {\sc goldfish} \citep{Shishkina2015, Kooij2018},
which applies a fourth-order finite-volume discretisation on staggered grids.
Three-dimensional direct numerical simulations (DNS) were performed for square cuboid domains with the aspect ratios $\Gamma = 2$, 2.5, 3, and 5.
The utilised staggered computational grids, which are clustered near all rigid walls,
are sufficiently fine to resolve the Kolmogorov microscales \citep{Shishkina2010}, see Tables~\ref{TAB} and \ref{TAB_par}.

\subsection{Experimental set-up}
\label{sec:exp_setup}

\begin{figure}
%figure 2
\centerline{\includegraphics[scale=1]{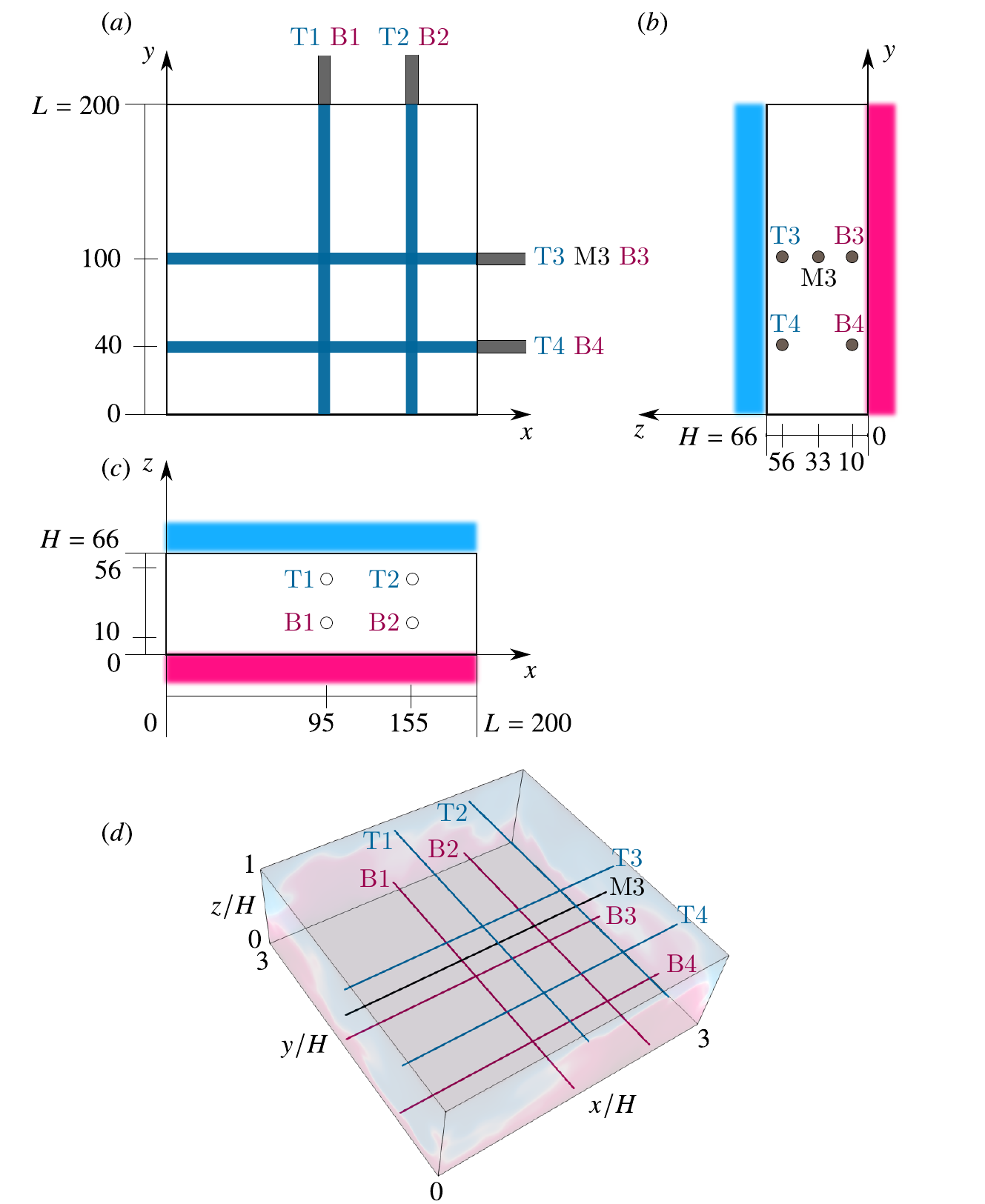}}
\caption{Schematics of the experimental set-up: (\textit{a--c)} three projections showing (\textit{a}) the top view and (\textit{b,c}) two side views and (\textit{d}) a three-dimensional sketch, illustrating the positions of all ultrasound transducers. 
Each ultrasound transducer is marked with a letter that indicates the distance to the bottom (``T'' -- close to the top, ``M'' -- matching the middle plane, ``B'' -- close to the bottom) followed by a number. 
All dimensional distances in (\textit{a--c}) are given in mm.
Blue and red colours indicate the cooled and heated plates respectively. 
}
\label{fig_setup}
\end{figure}

A schematic of the experimental set-up is presented in figure~\ref{fig_setup} along with the measuring positions of ultrasound probes. 
The set-up consists of a cuboid vessel with a base area of $L\times L=200 \times 200$ $\mathrm{mm^2}$ and a height $H=66$~mm, resulting in an aspect ratio $\Gamma \approx 3$. 
The top and bottom plates of this vessel are made of copper plates, whereas the side walls are made of polyvinyl chloride (PVC) of 30~mm thickness. 
This vessel is filled with an eutectic liquid metal alloy GaInSn of Gallium, Indium, and Tin, that serves as the working fluid in the experiment.
Thermophysical properties of GaInSn are reported in \citet{Plevachuk2014}.
In particular, the melting point of GaInSn is $10.5^\circ\text{C}$ and the Prandtl number equals $\Pran \approx 0.03$.

The liquid layer enclosed within the vessel is heated from the bottom and cooled from the top by adjusting the temperature of water flowing through channels in the copper plates. 
The temperature of water in these channels is held constant at set temperatures via two external thermostats. 
To minimise heat losses, the tubes transporting the hot and cold water and the entire vessel are wrapped in about 30~mm thick insulating foam tubes and additional envelope.
Two platinum resistance thermometers (Pt-100) (accuracy of $\pm 0.005$) have been utilised to accurately monitor the temperatures of water entering ($T_{in}$) and leaving ($T_{out}$) the hot and cold plates, respectively. 
These temperature readings are essential in measuring the non-dimensional convective heat transport, the Nusselt number $\NU$, expressed as $\NU=\dot{\Phi}/\dot{\Phi}_{cond}$. 
Here, $\dot{\Phi}_{cond}=\lambda L^2 \Delta / H$ is the conductive heat flux, with $\lambda$ being the thermal conductivity of the liquid metal. 
$\dot{\Phi}=\rho c_p \dot{V} (T_{in}-T_{out})$ is the total heat flux exchanged in the set-up, whereas $c_p$ is the isobaric heat capacity of water, and $\dot{V}$ is the flow rate of the circulating water determined via an axial turbine flow sensor at the cooling outlet of the set-up.

Prior to measurements, calibrations are performed. To account for the measurement uncertainty and heat losses of the set-up; hose split valves are used to split cold and hot water outlets respectively. One set of a cold and hot pair is used to feed the top plate and the other set to that of the bottom plate while ensuring that the temperature of both the plates remained at a set temperature of $20^\circ\text{C}$ using the external thermostats. Once the temperature in the plates reaches an equilibrium; an hour-long time series of temperature readings from both sets of thermocouples are recorded. Using the least square method, offsets from each of these thermocouples are extracted which are then used to correct the temperature measurements. This procedure gives a lower threshold of temperature difference attainable for the set-up, measurements below $\Delta \leq 0.22^\circ\text{C}$ are untenable. The range of measured temperature difference realized in this set-up varied from $0.27^\circ\text{C} \leq \Delta  \leq 16^\circ\text{C}$, with Rayleigh number in the range of $2.9 \times 10^4 \leq Ra \leq1.6 \times 10^6$. Experimental results presented here, see table~\ref{TAB_par}, are recorded after the temperature difference between the hot and the cold plates reached a constant value, when the system attains thermal equilibrium.

Principles of Ultrasound Doppler Velocimetry (UDV), a technique widely used for opaque flow diagnostics, are implemented to determine the fluid velocity \citep{Tsuji2005,Eckert2007}. 
Nine UDV transducers (TR0805SS, Signal Processing SA) are installed in a direct contact with the fluid. 
Each of these transducers acquire an instantaneous velocity profiles sequentially along the measuring lines as shown in figure~\ref{fig_setup} using multiplexing. The velocity measurements are performed with a resolution of about 0.5~mm/s and a sampling frequency of 1~Hz.

For the numerical results, statistical equilibrium or convergence is reached after several hundreds of free-fall time units. 
Throughout this paper, the length, velocity, and time are made non-dimensional using the cell height $H$, the free-fall velocity $u_{f\!f}$, and the free-fall time unit $t_{f\!f}\equiv H/u_{f\!f}$, respectively, see equations~(\ref{dimless}).    

\subsection{Phase averaging procedure}
\label{sec:phase_avg}

To analyse the 3D flow dynamics from the experimental data,
the whole field mapping of the velocity flow field is required, which is currently not possible using the UDV techniques. 
However, this sort of flow-field measurements can be assessed via the numerical techniques. 
The flow pattern consists of oscillatory coherent structures over several range of scales. 
To visualise the coherent structures, it is advisable to remove the background turbulent fluctuations using statistical means. 
\cite{Pandey2018} implemented an averaging method, which was later adopted by \cite{Akashi2022} in the form of a phase averaging algorithm. 
In this algorithm, one complete oscillation period, $\tau_{OS}= 1/f_{OS}$, is equally divided into certain (e.g. 16) intervals or phases.
Averaging of the temperature and velocity field data is carried out within each of these phases. 
This method reveals the underlying coherent structures in a flow field with high oscillations, such as that encountered in the three-dimensional cellular regime by \cite{Akashi2022}.

\cite{Vogt2018} used conditional averaging to showcase the 3D structures of the JRVs in a cylinder. The method of conditional averaging is similar to that of the phase-average process, with the only difference in the choice of the conditioning intervals. In the conditional averaging approach, the intervals for one complete cycle are divided into seven intervals bounded by multiples of standard deviations of the average temperature of the fluid. 

In the present study, the phase averaging method is applied to the simulation data, which cover 16 oscillation periods for the cases $\Gamma = 2.5$, $\Gamma = 3$, $\Gamma = 5$, and 8 oscillation periods for the case $\Gamma = 2$.
Every oscillation period is divided into 16 phases and each phase is represented by 20 snapshots of all flow fields.
Then the corresponding snapshots are averaged within each phase.
Finally, the conditional averaging is applied to the flow fields within each phase and for all oscillation periods, which gives a phase-averaged temporal evolution of all flow fields during the period.

\section{Results}

\label{sec:flow structure}

\textbf{\begin{table}
		\begin{center}
			\def~{\hphantom{0}}
			\small
			\begin{tabular}{llccccc}
				$\Gamma$ & &$\Pran$ &$\RA$ & $\ f_0 H^2/ \kappa \ $ & $\ f_0 (H+L')^2/ \kappa \ $ & $\ \NU\ $\\ \hline
				2   & DNS  & $0.03$ & $1.0 \times 10^6$ & 9.69 & 87.22 & 5.07 \\ \hline
				2.5 & DNS  & $0.03$ & $1.2 \times 10^5$ &  2.60 &  31.81 & 2.92 \\
					& DNS  &        & $1.0 \times 10^6$ &  7.21 & 88.32 & 5.31 \\ \hline
				3   & DNS  & $0.03$ & $1.0 \times 10^5$ &       &        & 2.92 \\
					& DNS  &        & $4.05 \times 10^5$ &  3.32 & 53.10 & 4.04 \\
					& DNS  &        & $1.0 \times 10^6$ &  5.22 & 83.52 & 5.18 \\ \hline
				3.03& Exp. & $0.03$ & $2.9 \times 10^4$ &       &        & 2.14 \\
					& Exp. &        & $3.5 \times 10^4$ &       &        & 2.37 \\
					& Exp. &        & $6.2 \times 10^4$ &       &        & 2.67 \\
					& Exp. &        & $6.4 \times 10^4$ &       &        & 2.55 \\
					& Exp. &        & $6.8 \times 10^4$ &       &        & 2.73 \\
					& Exp. &        & $6.9 \times 10^4$ &       &        & 2.71 \\
					& Exp. &        & $9.4 \times 10^4$ &       &        & 2.96 \\
					& Exp. &        & $1.0 \times 10^5$ &       &        & 2.95 \\
					& Exp. &        & $1.1 \times 10^5$ &       &        & 3.02 \\
					& Exp. &        & $1.2 \times 10^5$ &       &        & 3.12 \\
					& Exp. &        & $1.6 \times 10^5$ &       &        & 3.37 \\
					& Exp. &        & $2.7 \times 10^5$ &       &        & 3.60 \\
					& Exp. &        & $3.2 \times 10^5$ &  3.62 & 58.81  & 3.70 \\
					& Exp. &        & $4.1 \times 10^5$ &  4.61 & 74.93  & 3.89 \\
					& Exp. &        & $5.1 \times 10^5$ &  5.09 & 82.61  & 4.19 \\
					& Exp. &        & $6.3 \times 10^5$ &  5.73 & 93.13  & 4.43 \\
					& Exp. &        & $7.7 \times 10^5$ &  6.21 & 100.81  & 4.61 \\
					& Exp. &        & $8.6 \times 10^5$ &  6.56 & 106.54 & 4.79 \\
					& Exp. &        & $9.4 \times 10^5$ &  6.72 & 109.18 & 4.85 \\
					& Exp. &        & $1.0 \times 10^6$ &  7.22 & 117.24 & 4.99 \\
					& Exp. &        & $1.2 \times 10^6$ &  7.57 & 123.00 & 5.18 \\
					& Exp. &        & $1.3 \times 10^6$ &  8.02 & 130.25 & 5.27 \\
					& Exp. &        & $1.6 \times 10^6$ &  8.53 & 138.60 & 5.51 \\ \hline
				5   & DNS  & $0.03$ & $1.2 \times 10^5$ &  3.21 &  39.33 & 3.04 \\
					& DNS  &        & $1.0 \times 10^6$ &  8.65 & 105.95 & 5.40 \\
			\end{tabular}
			\caption{Details on the conducted DNS and experiments.}
			\label{TAB_par}
		\end{center}
\end{table}}

\begin{figure}
%figure 3
\centerline{\includegraphics[scale=1]{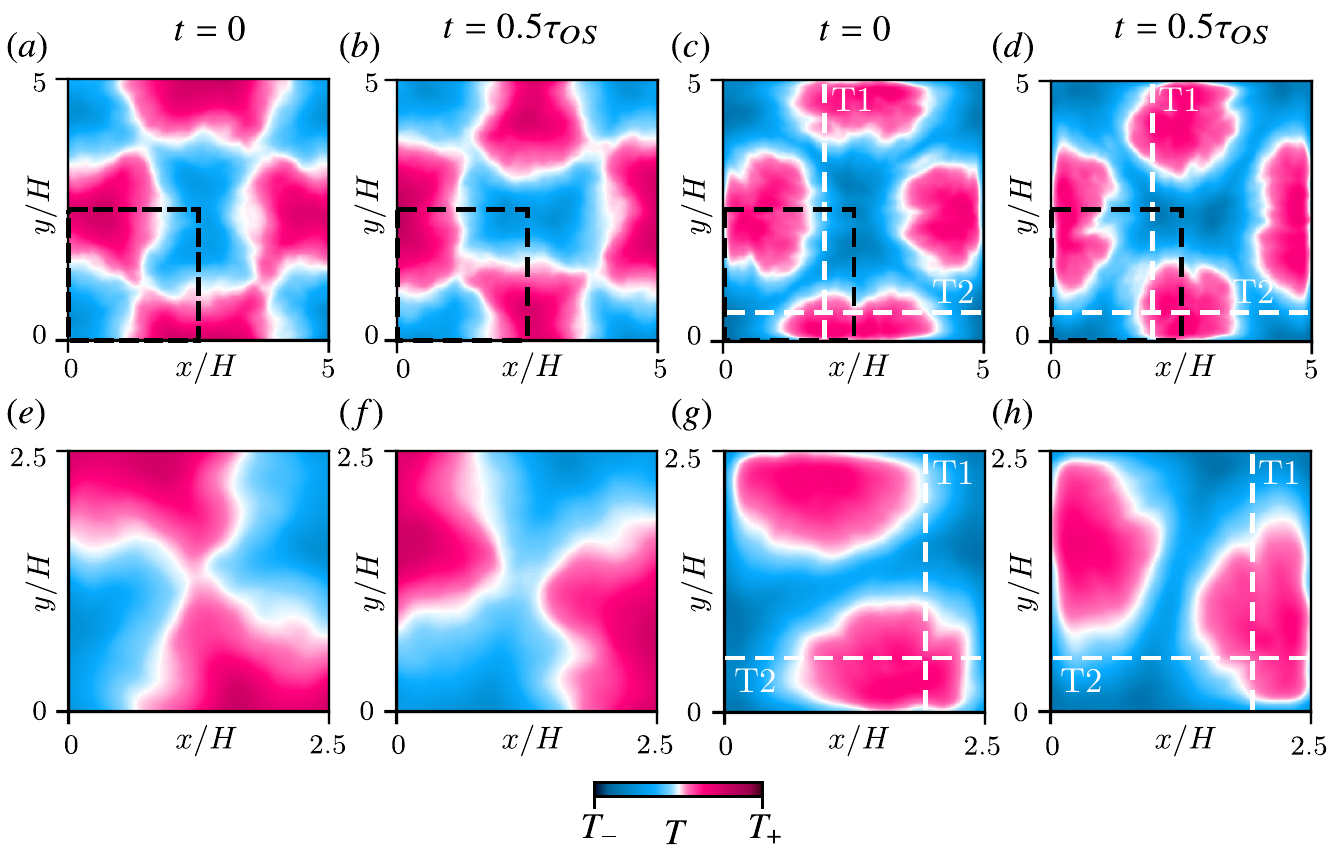}}
\caption{Phase-averaged snapshots of the temperature at a distance $z$ from the bottom: (\textit{a,~b,~e,~f}) $z= 0.5H$ and (\textit{c,~d,~g,~h}) $z=0.85H$, for different container aspect ratios (\textit{a--d}) $\Gamma = 5$ and (\textit{e--h}) $\Gamma = 2.5$, as obtained in the simulations for $\RA = 10^6$ and $\Pran = 0.03$ (see supplementary movies). 
The virtual probe lines T1 and T2 (see also figure~\ref{sim_G5_vs_G2p5}) are indicated with dashed white lines.
The black squares indicate the areas that correspond to the areas of the container with $\Gamma=2.5$.
}
\label{T_slices}
\end{figure}

The results of all conducted DNS and experiments are summarised in Table~\ref{TAB_par}.
For all experimental and numerical data for sufficiently large $\RA$, an oscillatory behaviour of the LSC was identified.
Like in the case of $\Gamma=\sqrt{2}$ and $\Gamma=2$ of a cylindrical container \citep{Vogt2018}, the JRV-like oscillatory structures leave imprints on almost all flow characteristics, for the considered ranges of $\RA$ and $\Gamma$ of a cuboid domain.
The oscillatory behaviour of the LSC is reflected in temporal evolution of the temperature and particular components of the velocity fields and is also seen in the vertical heat flux temporal evolution.

Once the dominating frequency $f_0$ is evaluated (we will discuss this in more detail later), one can analyse the mean flow dynamics within the time period that lasts $\tau_{OS}= 1/f_0$.
For that, the temporal evolution of the flow fields, which are obtained in the DNS, are split into separate periods, according to the dominating frequency $f_0$, and then a phase-averaged temporal evolution of all flow fields during the period is calculated.

Our DNS for $\RA = 10^6$ and $\Pran = 0.03$ and two different aspect ratios, $\Gamma=5$ and $\Gamma=2.5$, show a very remarkable similarity of the global flow structure and its dynamics.
 In figure~\ref{T_slices}, phase-averaged instantaneous temperature distributions in horizontal cross-sections are presented, which are considered at  distances $z= 0.5H$ (figure~\ref{T_slices}~\textit{a,~b} and \textit{e,~f}) and $z= 0.85H$ (figure~\ref{T_slices}~\textit{c,~d} and \textit{g,~h}) from the bottom plate, and at the times $t=0$ and $t=0.5\tau_{OS}$.
This figure shows patches of upwelling (hot) and downwelling (cold) fluid with the hot patches connected by a diagonal ridge of upwelling fluid. These patches rotate counterclockwise in the time interval $[0,\,0.5\tau_{OS}]$ (see supplementary movies), suggesting the presence of oscillatory flow dynamics which periodically changes the flow topology. 
For fixed values of $\RA$ and the cell height $H$, the spatial length of the convection cell in the case $\Gamma=5$ is twice larger than in the case $\Gamma=2.5$ for the same $\RA$ and $H$. 
Therefore, for any fixed $z$, one can expect a similarity of the flow pattern in the horizontal cross-section at the height $z$ in the case $\Gamma=2.5$ with the flow pattern in the 1/4 of the area of the horizontal cross-section at the same height $z$ in the case $\Gamma=5$.
Indeed, figure~\ref{T_slices} shows that the temperature distribution in the region marked with black dashed lines for $\Gamma=5$  (figure~\ref{T_slices}~\textit{a--d}) is very similar to the temperature distribution in the corresponding cross-sections for $\Gamma=2.5$  (figure~\ref{T_slices}~\textit{e--h}) if considered at the same phase.

\begin{figure}
%figure 4
\centerline{\includegraphics[scale=1]{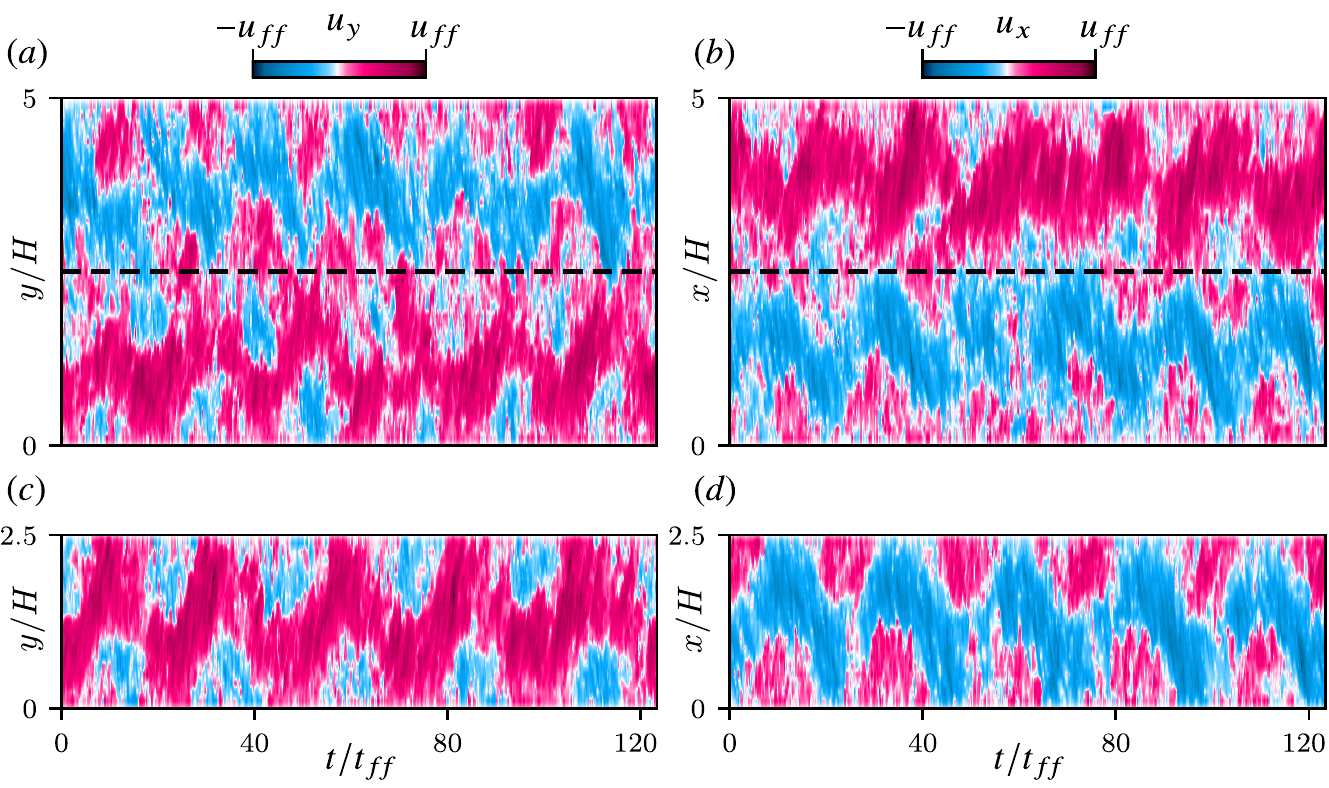}}
\caption{Spatio-temporal velocity maps for $\RA = 10^6$ at $z= 0.85H$, as obtained in the direct numerical simulations at the virtual probe lines T1~(\textit{a,~c}) and T2~(\textit{b,~d}), for the aspect ratios $\Gamma=5$~(\textit{a,~b}) and $\Gamma = 2.5$~(\textit{c,~d}).
The black dashed lines in (\textit{a,~b}) correspond to the measurements in the cuboid with $\Gamma=2.5$ (\textit{c,~d}), respectively.}
\label{sim_G5_vs_G2p5}
\end{figure}

To gain more evidence for this similarity, we evaluate the horizontal components of the velocity, $u_y$ and $u_x$, along the lines marked T1 and T2 in figure~\ref{T_slices}~(\textit{c,~d}) ($\Gamma=5$) and compare them with the corresponding horizontal components of the velocity along the lines marked T1 and T2 in figure~\ref{T_slices}~(\textit{g,~h}) ($\Gamma=2.5$).
The temporal evolutions of these velocity components for $\Gamma=5$ and $\Gamma=2.5$ are compared in figure~\ref{sim_G5_vs_G2p5} for $\RA = 10^6$ and $z= 0.85H$.
One can see that the lower halves of the spatio-temporal velocity maps in  figure~\ref{sim_G5_vs_G2p5}~(\textit{a,~b}), which correspond to the measurements along the lines T1 and T2 within the 1/4 area that is  marked in figure~\ref{T_slices}~(\textit{c,~d}) with the black dashed lines, mimic the spatio-temporal velocity maps in  figure~\ref{sim_G5_vs_G2p5}~(\textit{c,~d}), which correspond to the measurements along the lines T1 and T2  in figure~\ref{T_slices}~(\textit{g,~h}).

Qualitatively, the signals for $\Gamma = 5$ and $\Gamma = 2.5$ are similar, however, the frequency of the oscillations in the latter case is slightly lower than that in the former case, with six versus five oscillations during the same time interval.
Also at $\Gamma = 5$ the signal seems to be less stable than in the case $\Gamma = 2.5$.

\begin{figure}
%figure 5
\centerline{\includegraphics[scale=1]{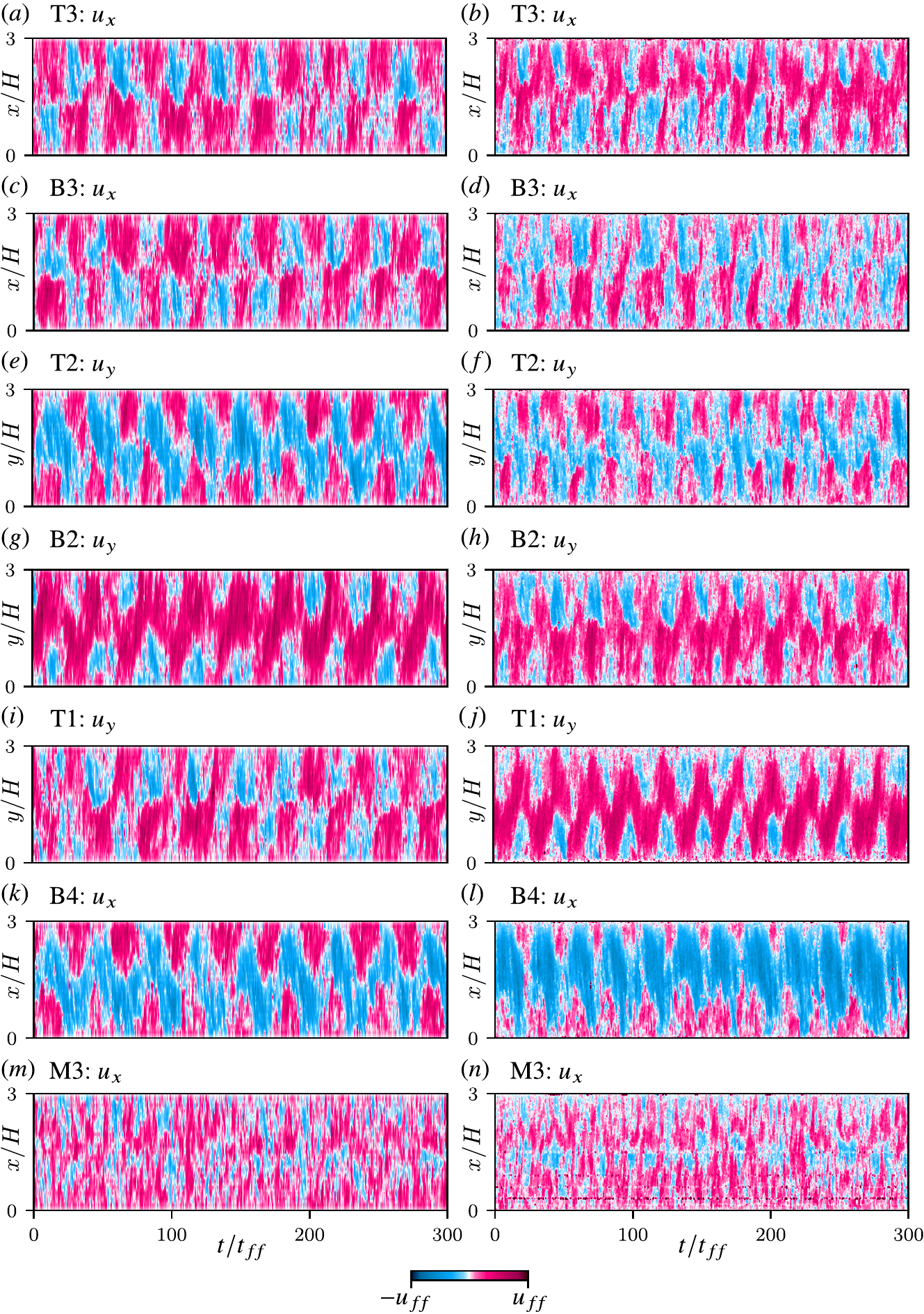}}
\caption{Spatio-temporal velocity maps for $\RA = 10^6$, as obtained in the direct numerical simulations (left column) 
and in the experimental measurements (right column), for the aspect ratio $\Gamma= 3$.
The numerical data are probed exactly at the same locations where the UDV sensors are located in the experiment: T3 (\textit{a,~b}), B3 (\textit{c,~d}), T2 (\textit{e,~f}), B2 (\textit{g,~h}), T1 (\textit{i,~j}), B4 (\textit{k,~l}) and M3 (\textit{m,~n}).
}
\label{num_exp}
\end{figure}

Figure~\ref{num_exp} shows a comparison of the experimental and simulation results for the same $\Gamma=3$ and $\RA=10^6$.
Here a comparison of the temporal evolution of the horizontal components of the velocity is made exactly at the same locations in the DNS and in the experiment.
One can see a good qualitative agreement between the experimental and simulation data.
However, the dominant frequency obtained in the experiment is slightly higher compared to the frequency evaluated from the simulation data.
More precisely, we obtained on average 11 oscillations in the experiment versus 9 oscillations in the DNS for the same time interval.

\subsection{Three-dimensional cellular flow dynamics}

\begin{figure}
%figure 6
\centerline{\includegraphics[scale=1]{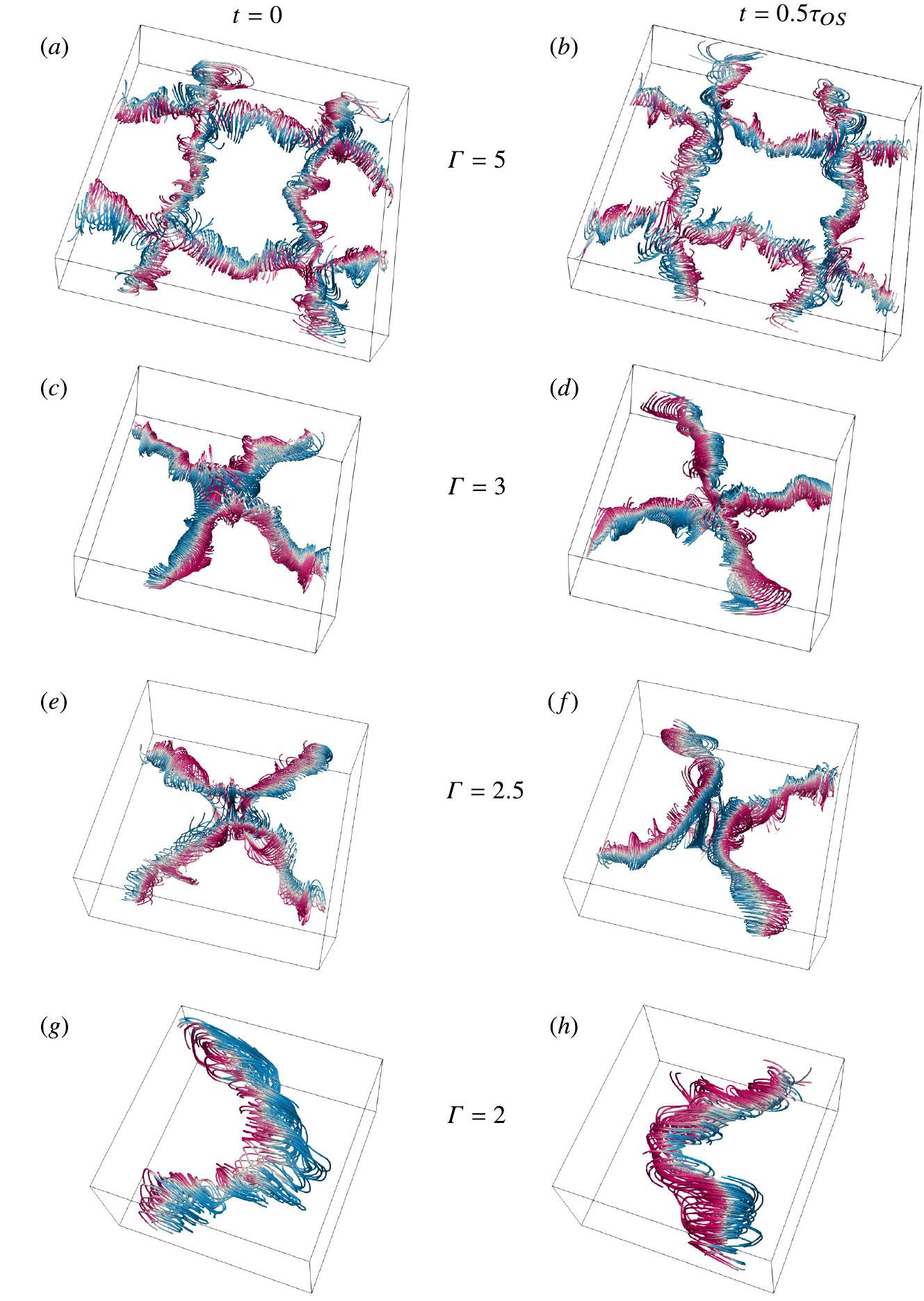}}
\caption{Phase-averaged streamlines in Rayleigh--B\'enard convection for $\Pran = 0.03$, $\RA = 10^6$, as obtained in the direct numerical simulations for parallelepiped domains with different aspect ratios:
$\Gamma = 5$~(\textit{a,~b}), $\Gamma = 3$~(\textit{c,~d}), $\Gamma = 2.5$~(\textit{e,~f}) and $\Gamma = 2$~(\textit{g,~h}).
Blue (red) colour corresponds to a negative (positive) value of the vertical velocity component $u_z$.
}
\label{streamline_vortex}
\end{figure}

Figure~\ref{streamline_vortex} shows phase-averaged streamlines in Rayleigh--B\'enard convection for $\Pran=0.03$, $\RA=10^6$, as obtained in the  direct numerical simulations for cuboid domains for all considered aspect ratios $\Gamma$ at the beginning ($t=0$) and at the middle ($t=0.5\tau_{OS}$) of the oscillation period.
There are four interlacing JRVs in the case $\Gamma = 5$ (figure~\ref{streamline_vortex}~\textit{a,~b}). 
The flow structure resembles a cellular structure which previously was observed in \cite{Akashi2022} for $\Gamma = 5$ and $\RA \approx 1.2 \times 10^5$.
There are only two JRVs for the aspect ratios $\Gamma = 3$ (figure~\ref{streamline_vortex}~\textit{c,~d}), and $\Gamma = 2.5$ (figure~\ref{streamline_vortex}~\textit{e,~f}), which is in contrast to a lattice of four JRVs in the $\Gamma=5$ case.
What is more striking is that the JRVs in the $\Gamma = 3$ and 2.5 cells represent a quadrant of the JRV lattice of the $\Gamma= 5$ cell (see also figure~\ref{JRV_family}).
Only one vortex is observed in the convection cell with $\Gamma = 2$.
This dynamic interplay between changing aspect ratio and organisation of the JRVs highlights the influence of shape and size of the geometry of the container. 
It also rises an important question as to whether there is a certain hierarchy in these systems when it comes to the reorganisation of JRVs within a container.

\begin{figure}
%figure 7
\centerline{\includegraphics[scale=1]{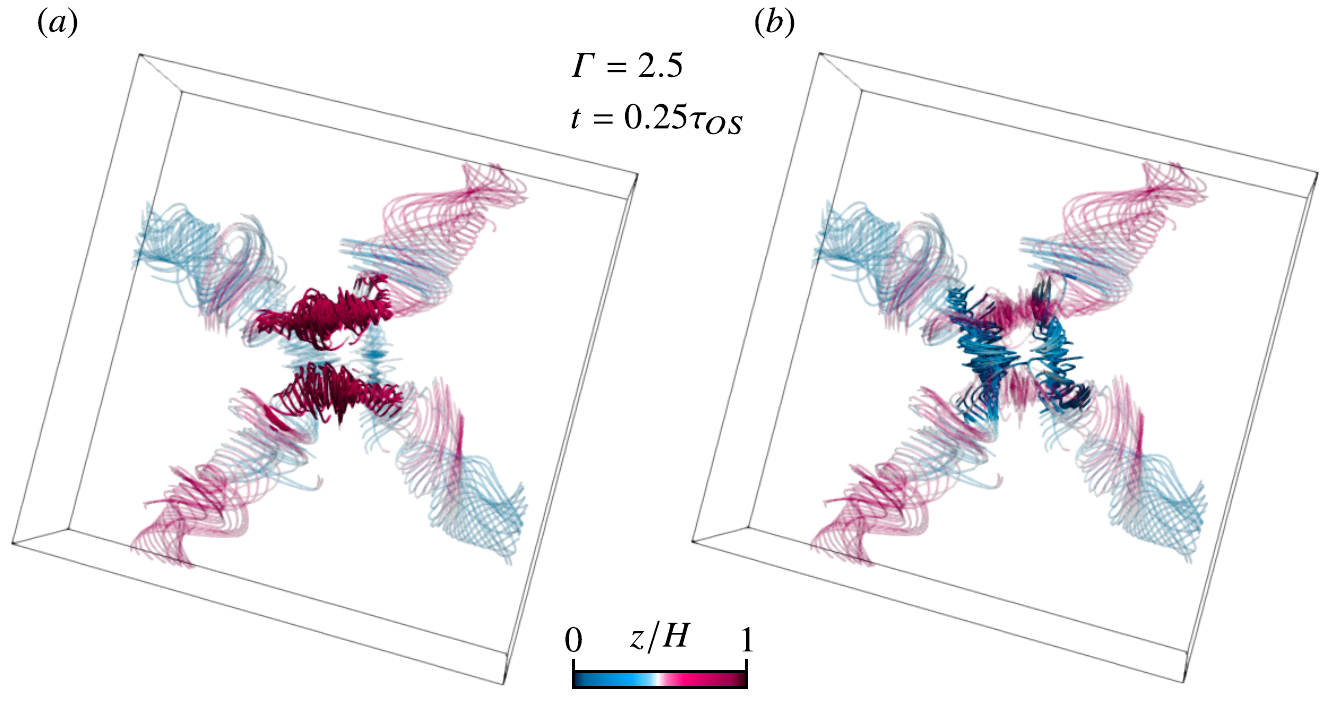}}
\caption{
Phase-averaged streamlines in Rayleigh--B\'enard convection for $\Pran=0.03$, $\RA=10^6$, as obtained in the direct numerical simulations for $\Gamma = 2.5$, $t = 0.25 \tau_{OS}$.
Vortices in the upper part of the domain are highlighted in (\textit{a}), vortices in the lower part of the domain are highlighted in (\textit{b}).
For convenience, all other streamlines outside the center of the domain are shown transparent.
Colours correspond to the vertical coordinate of the structure $z$.
}
\label{streamline_vortex2}
\end{figure}

A closer look at the 3D flow structure for $\Gamma= 2.5$ shows how the two vortices connect to each other in the central part of the box (see supplementary movies). 
In this case, the JRVs are connected with two vortices in the upper part and two vortices in the lower part of the domain.
Figure~\ref{streamline_vortex2} shows the phase-averaged streamlines for the same $t=0.25\tau_{OS}$, but the connecting vortices in the upper central part of the domain are highlighted with red colour in figure~\ref{streamline_vortex2}~\textit{a}, while the connecting vortices in the lower part of the domain are highlighted with blue colour in figure~\ref{streamline_vortex2}~(\textit{b}). 
Thus colours here reflect the distance $z$ from the bottom plate, i.e., the vertical coordinate of the structure.
For clarity, all other streamlines are shown transparently. 
It is also worth noting that for the same $\RA$, the JRVs are more stable and better pronounced for $\Gamma=2.5$ compared to $\Gamma=3$.

\begin{figure}
%figure 8
\centerline{\includegraphics[scale=1]{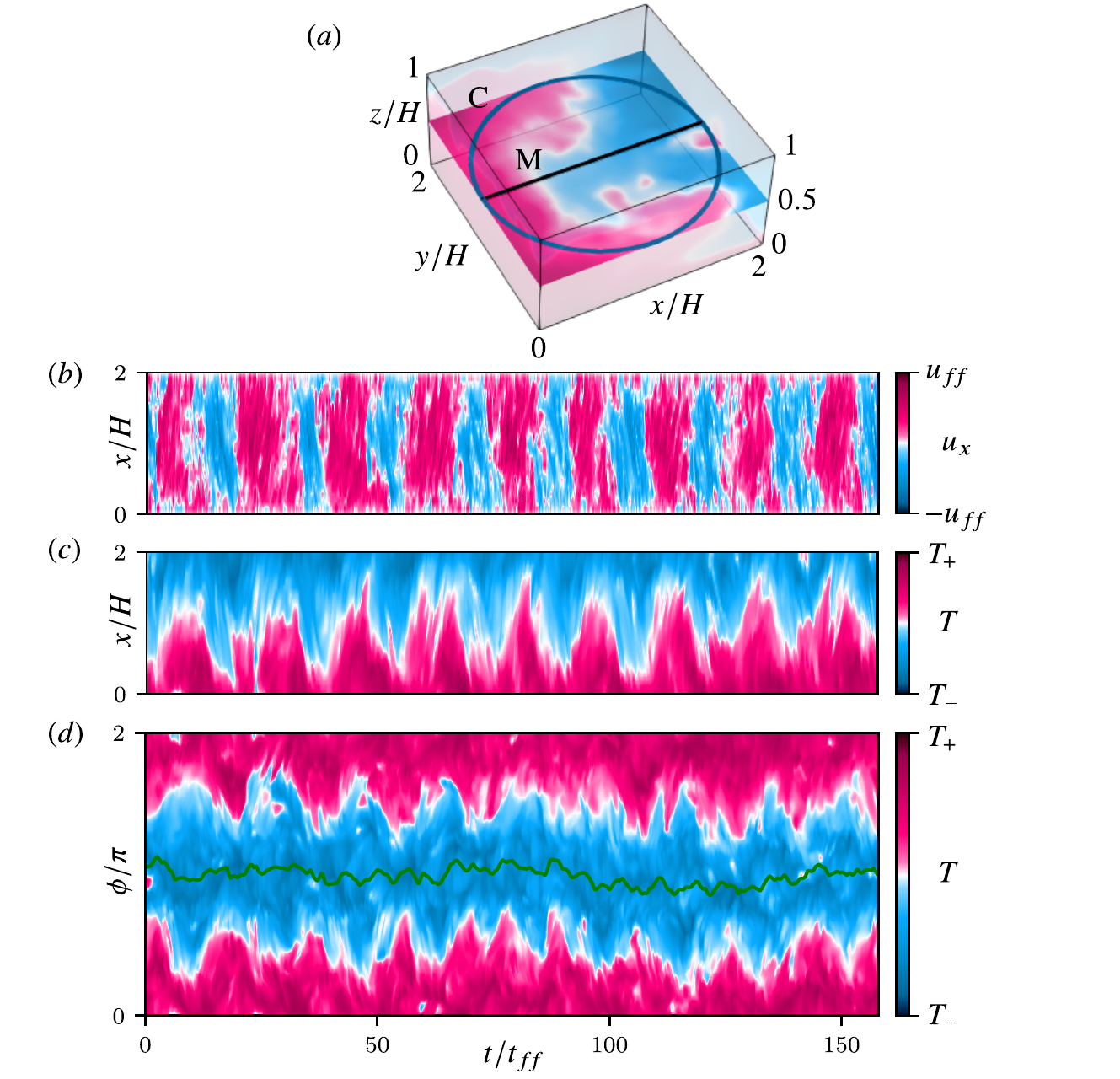}}
\caption{Data for  $\RA = 10^6$ for $\Gamma = 2$, as obtained in the direct numerical simulations. 
Measurement positions in the midplane at $z = 0.5 H$  are shown in (\textit{a}): the central straight line marked ``M'' and the circle marked ``C'' are shown in black and blue colours, respectively.
Spatio-temporal (\textit{b}) velocity and (\textit{c}) temperature maps along the M-line.
(\textit{d}) Spatio-temporal temperature map along the C-circle.
The instantaneous position angle of the LSC is marked with the green line (cf. figure~4 in~\cite{Vogt2018}, for a cylinder with the same $\Gamma$ and $\RA$). 
}
\label{G2_time_evol}
\end{figure}

Figure~\ref{G2_time_evol} shows in detail the specifics of the spatial-temporal velocity and temperature maps for $\Gamma = 2$.
In this case (see figure~\ref{streamline_vortex}~\textit{e,~f}) there is only one vortex that rotates in the direction opposite to the LSC direction.
Note that for the considered $\RA = 10^6$, the LSC is oriented along a vertical wall of the container rather than diagonally.
The flow pattern is similar to that obtained for the $\Gamma = 2$ cylinder~\citep{Vogt2018}.
To examine this similarity, we evaluate at the mid-height, at $z = 0.5 H$ (figure~\ref{G2_time_evol}\textit{a}), the horizontal component of the velocity $u_x$ and the temperature along the straight line marked in the figure ``M'' and along the circle marked ``C'', which correspond to the lines along the diameter and along the midplane circumference of an inscribed cylinder \citep[as it was considered in][]{Vogt2018}, respectively.

The dominant frequency $f_0$ is visible in the spatio-temporal maps of both velocity (figure~\ref{G2_time_evol}\textit{b}) and temperature (figure~\ref{G2_time_evol}\textit{c}).
Figure~\ref{G2_time_evol}\textit{b} resembles figure 2~\textit{c} in\cite{Vogt2018}.

\begin{figure}
%figure 9
\centerline{\includegraphics[scale=1]{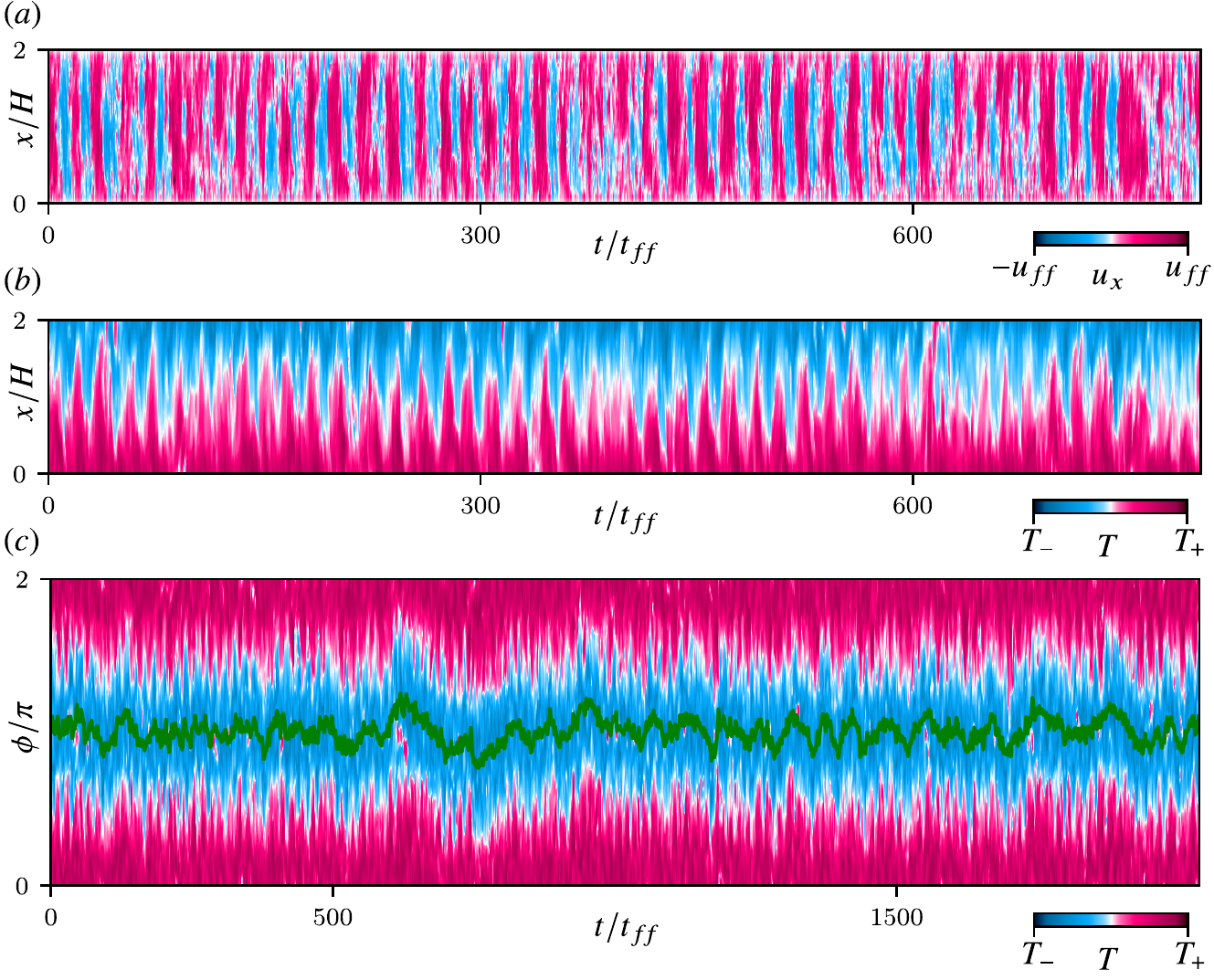}}
\caption{Data for $\RA = 10^6$ and $\Gamma = 2$, as obtained in the direct numerical simulations. 
A similar figure to figure~\ref{G2_time_evol}~(\textit{b-d}), but the time series is longer.
Spatio-temporal (\textit{a}) velocity and (\textit{b}) temperature maps along the M-line.
(\textit{c}) Spatio-temporal temperature map along the C-circle.
The instantaneous position angle of the LSC is marked with the green line.
}
\label{G2_time_evol_long}
\end{figure}

The spatio-temporal maps of the temperature along the circle ``C'' (figure~\ref{G2_time_evol}\textit{d}) also look similar to those in figure 4(\textit{a}) in~\cite{Vogt2018} and indicate the presence of a dominant frequency.
It is worth noting that the oscillations of the LSC orientation, which are characterised by the azimuthal angle $\xi_{LSC}$, and computed here using the single-sinusoidal fitting method by~\cite{Cioni1997} (indicated with the green line in figure~\ref{G2_time_evol}\textit{d}) are strong. Although, in cuboid domain, the LSC direction is expected to be more stable compared to that of a cylindrical domain.
This is clearly demonstrated by longer time series, which are presented in figure~\ref{G2_time_evol_long}.
In contrast to the relatively short time interval with only 9 oscillation periods, when the LSC orientation was mostly stable (figure~\ref{G2_time_evol}), the longer time series reveal quite strong temporal oscillations of the azimuthal angle $\xi_{LSC}$ (figure~\ref{G2_time_evol_long}\textit{c}).
Spatio-temporal maps for both velocity and temperature (figure~\ref{G2_time_evol_long}~\textit{a,~b}) show that intervals with relatively stable regular  oscillations alternate with intervals with less stable signals.
This leads to difficulties for conditional averaging in this case.
Therefore, as mentioned in section \ref{sec:phase_avg}, we used averaging for only 8 oscillation cycles for $\Gamma=2$, whereas for other values of $\Gamma$ the averaging was performed for 16 oscillation cycles.

\subsection{Oscillation frequency}
Any periodic oscillation of the JRV has a certain dominant frequency $f_0$. 
For each studied case, these frequencies were extracted both from the velocity and temperature time series.
The frequencies were non-dimensionalised using the dissipative time scale. 
Two length scales were used for the dissipative time scale: the height of the domain $H$ and the overall path length of the LSC, following \cite{Cheng2022}, is coarsely approximated as $2l$, $l\equiv H+L'$, where $L'$ equals to $2H$, $2.5H$, $3H$ and $5/2H$ for  $\Gamma = 2$, 2.5, 3 and 5 respectively.
Note that in the latter case $L'= 5/2H$, since for the $\Gamma = 5$ container, the flow pattern consists of two JRV building blocks, repeated in both horizontal directions, as it was shown above.
The diffusion times are then $\tau_\kappa = H^2/\kappa$ and $\tau^l_\kappa = (H+L')^2/\kappa$ with the corresponding diffusion frequencies $f_k = 1/\tau_\kappa = \kappa/H^2$ and $f^l_k = 1/\tau^l_\kappa = \kappa/(H+L')^2$ respectively.

\begin{figure}
%figure 10
%\unitlength1truecm
\centerline{\includegraphics[scale = 1]{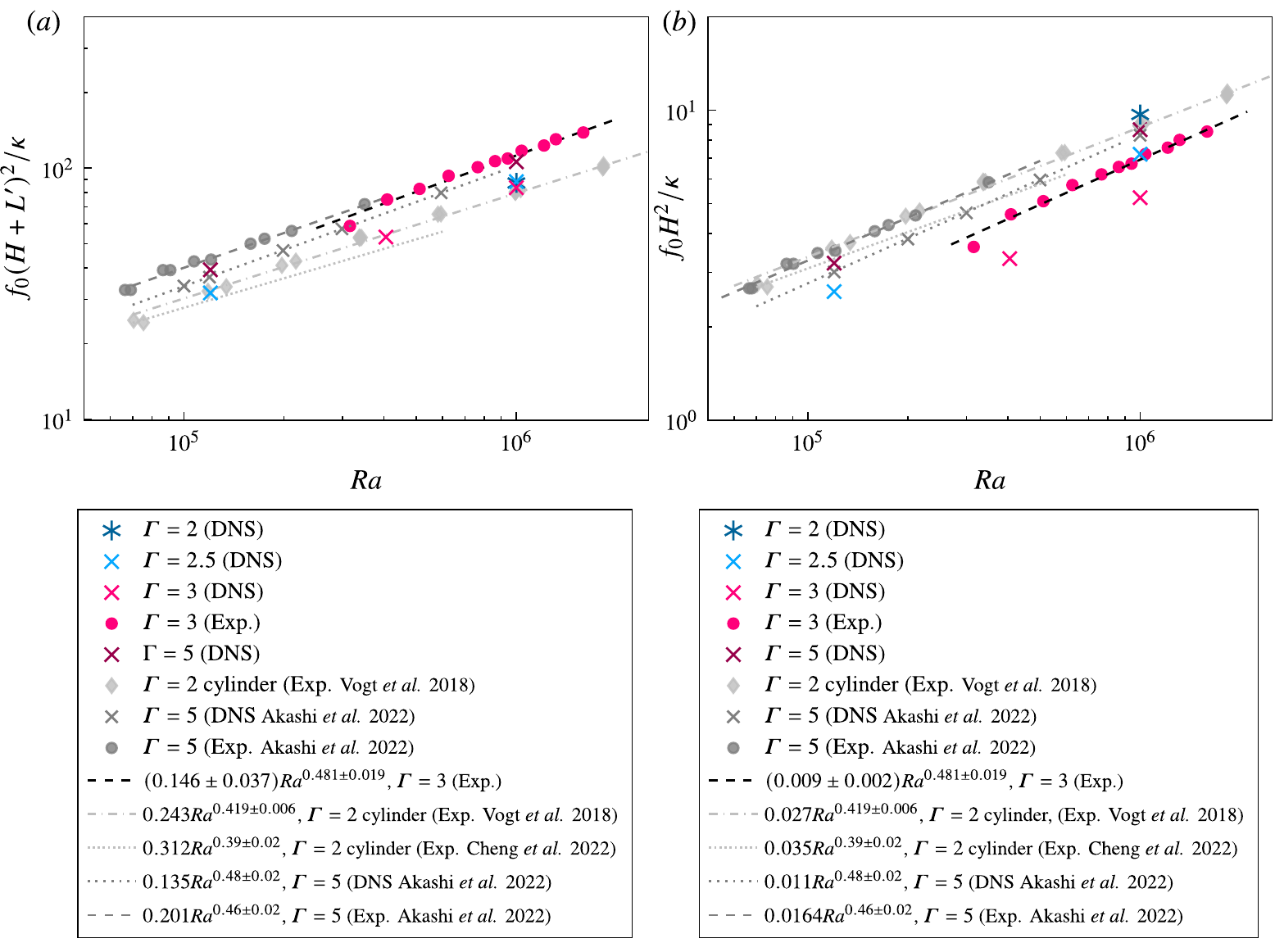}}
\caption{Dominant frequencies $f_0$, which are non-dimensionalised using the dissipative time scales (\textit{a}) $(H+L')^2/\kappa$ and (\textit{b}) $H^2/\kappa$, as functions of $\RA$. 
Here $L'$ equals to $2H$, $2.5H$, $3H$ and $5/2H$ for  $\Gamma = 2$, 2.5, 3 and 5, respectively.}
\label{fig_frequency}
\end{figure}

Figure~\ref{fig_frequency} shows the values of the dominant frequencies $f_0$ versus $\RA$ for all considered values of $\Gamma$.
The data from the present study are shown in red and blue colours, all other data are shown in grey colour.
The values that the frequencies take in different flow configurations are presented in table~\ref{TAB_par}.
Variation of $f_0$, normalised with the dissipative time scale $(H+L')^2/\kappa$, is shown in figure~\ref{fig_frequency}\textit{a}, as a function of $\RA$.
One can see that our new experimental data for $\Gamma = 3$ (red circles) and those from \cite{Akashi2022} for $\Gamma = 5$ (grey circles) collapse onto one master scaling line.
Note that for $\Gamma = 3$ the oscillatory JRV mode occurs at higher $\RA$ compared to the case of $\Gamma = 5$.
The so-called roll regime at lower $\RA$ does not have clear dominant frequencies, therefore only data for the cases where an oscillatory mode exists are presented in the figure.
A comparison between the experimental and simulation results for $\Gamma = 3$ shows that the oscillation frequency values obtained in the experiment (as it was already shown in figure~\ref{num_exp}) is slightly higher than that evaluated from the simulation data for $\Gamma = 3$.
%%%%%%%However, one should take note that the values of $f_0$ obtained in the simulations by \cite{Akashi2022} were also systematically lower compared to the experimental data.
%

The numerically obtained normalized frequencies for $\Gamma = 3$, 2.5, and 2 are very close to each other.
For $\Gamma = 5$, which is the only case with two JRV building blocks, the normalized frequency is generally higher than the frequencies for other $\Gamma$ (cf. crosses and asterisk at $\RA = 10^6$).
The numerically obtained dimensionless frequency for $\Gamma=5$ at $\RA = 10^6$ is in very good agreement with the scaling line for $\Gamma=5$ reported in \cite{Akashi2022} and with the new experimental data for $\Gamma=3$.
For a lower Rayleigh number, $\RA = 1.2 \times 10^5$, the numerically obtained dimensionless frequency for $\Gamma=5$ is also in very good agreement with the numerical and experimental data from \cite{Akashi2022}.
Experimental data for a $\Gamma = 2$ cylinder from \cite{Vogt2018}
%with $\Pran = 0.025$ 
and \cite{Cheng2022} 
%with $\Pran = 0.02$ 
give scaling relations with slightly lower exponent values and for the considered $\RA$ range; they locate slightly below the fitting lines for $\Gamma = 3$ and 5.
The frequency value from our $\Gamma = 2$ box simulations is close to that for $\Gamma = 2$ cylinder experimental data from \cite{Vogt2018}.  
The numerical data for all considered $\Gamma$ are located between the fitting lines obtained in the experiments for $\Gamma = 3$, 5 box and cylinder $\Gamma = 2$.

To sum up, all the experimental and numerical data show a very similar frequency dependence, as one can see in a $f_0$ versus $\RA$ plot, across all aspect ratios with the frequency normalisation based on the path length $l$.
In figure~\ref{fig_frequency}\textit{b} we normalise the frequency $f_0$ with the thermal diffusion time $\tau_\kappa = H^2/\kappa$.
In that case, without taking into account the spatial length of the vortex path, the deviation between the data points for $\Gamma = 5$ and 2.5 remains the same (as the length scale is the same for these two cases), while the data points for $\Gamma = 3$ move down and the points for $\Gamma = 2$ move significantly up.
We conclude that the spatial length of the domain is an important control parameter, which together with the height of the fluid layer determines the relevant length and the scaling relations for the oscillation frequency.

\subsection{Heat transport}

In this section, the effect of the flow dynamics on the heat transport is discussed.
The volume-averaged Nusselt number $\NU_{vol}$ can be evaluated from the simulation data as follows:
\begin{eqnarray}
\NU_{vol}=\langle \Omega_z \rangle_{V,t},
\label{NuOB}
\end{eqnarray}
 where $\Omega_z$ is a component of the full heat flux vector $\vec{\Omega} \equiv (\vec{u}T - \kappa \vec{\nabla} T) / (\kappa \Delta /H)$ along the vertical axis and $\langle \cdot\rangle_{V,t}$ denotes the the time--volume average.
In the experiments, the  Nusselt numbers $\NU$ are computed as discussed in section \ref{sec:exp_setup}.

\begin{figure}
%figure 11
\centerline{\includegraphics[scale = 1]{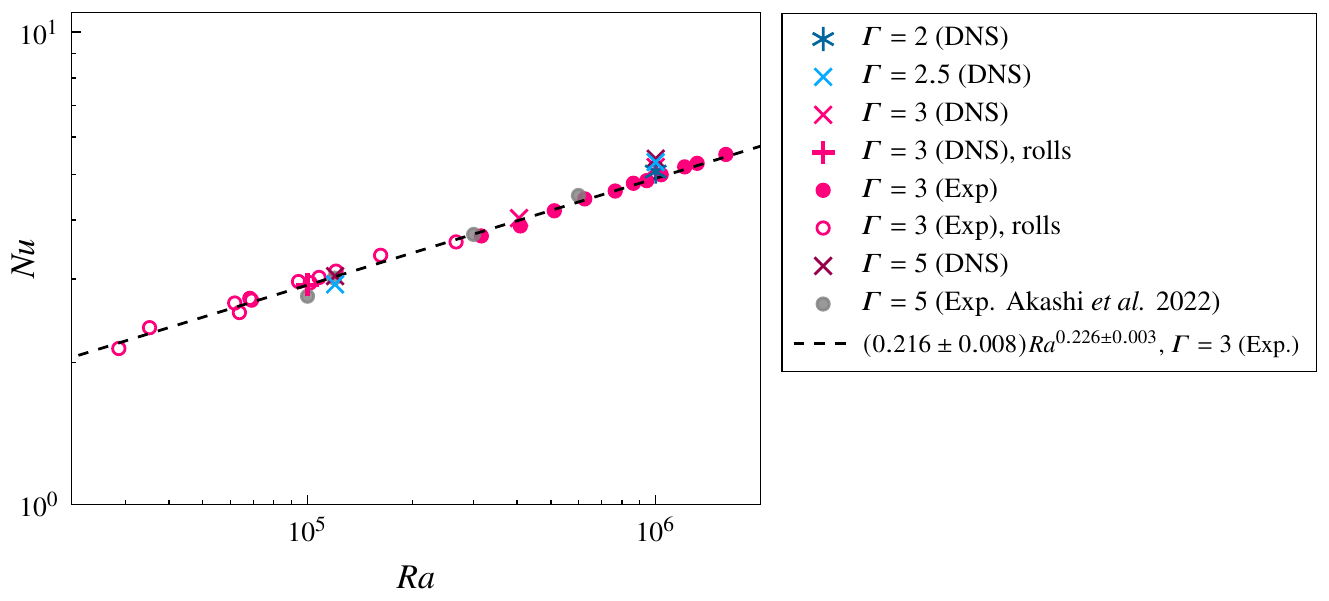}}
\caption{Scaling of the Nusselt number $\NU$ with the Rayleigh number $\RA$, for all studied $\Gamma$.
%An increase of the $\RA$ changes the global flow structure from two rolls to 3D cellular regime. 
}
\label{fig_ranu}
\end{figure}

The global heat transport scaling across various $\RA$ is shown in figure \ref{fig_ranu}.
The flow dynamics does not seem to have any dramatic effect on the heat transport. 
This is true for all studied aspect ratios. 
The cases without oscillations are shown in the figure with open symbols.
The fitted curve gives a scaling relation $\NU = 0.22 \times \RA^{0.23}$, which differs slightly from that reported by \cite{Vogt2021}:  $\NU= 0.166 \times \RA^{0.25}$. 
However, this difference can be attributed to the difference in the geometry of the cell.

\begin{figure}
%figure 12
\centerline{\includegraphics[scale=1]{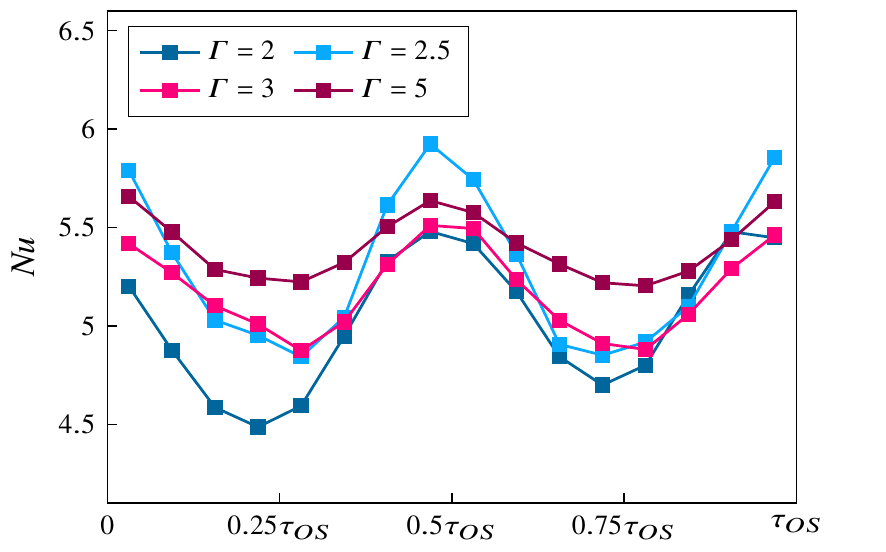}}
\caption{Phase-averaged Nusselt number $\NU (t)$ during one oscillation period, as it is evaluated from the direct numerical simulations for $\RA = 10^6$ and different aspect ratios $\Gamma$ of a parallelepiped container.}
\label{fig_phaseavgheat}
\end{figure}

An interesting feature of the JRV regimes is that the Nusselt numbers, which are computed using the phase-average method as discussed in section \ref{sec:phase_avg}, demonstrate an oscillatory behaviour during the JRV cycle.
This is demonstrated for one period of oscillation in figure \ref{fig_phaseavgheat}.
Qualitatively, the oscillatory behaviour of the local vertical heat flux $\NU(t)$ during the JRV cycle is the same for all considered $\Gamma$.
It is clear that the $\NU$ shows oscillatory behaviour with distinct peaks of maxima and minima. 
This sort of behaviour was also reported in previous study of \cite{Akashi2022} for $\Gamma = 5$.
However, the amplitude of the oscillations for a certain given $\RA$ is different: it decreases with increasing $\Gamma$.

\begin{figure}
%figure 13
\centerline{\includegraphics[scale=1]{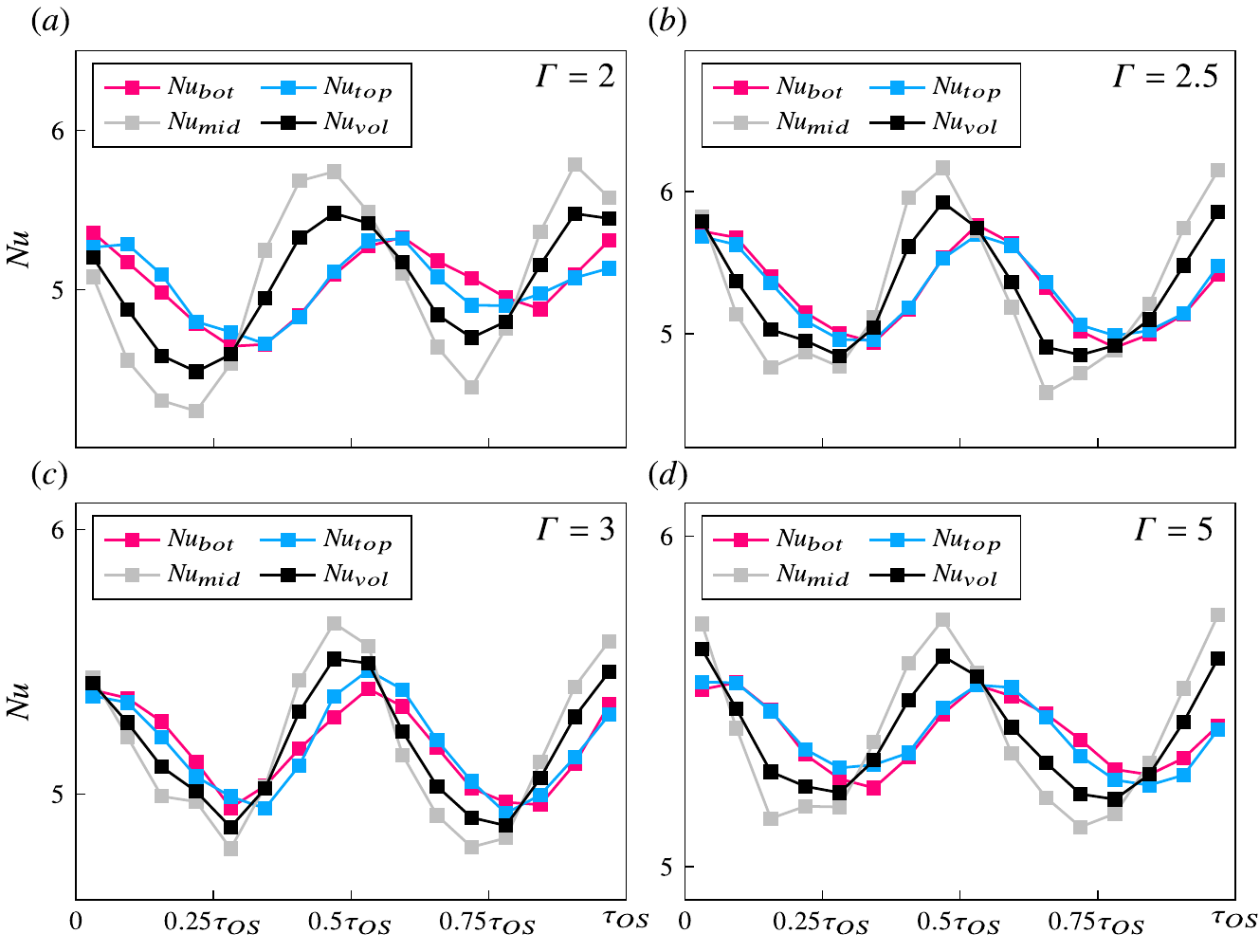}}
\caption{Evolution of the phase-averaged Nusselt number $\NU (t)$ during the one oscillation period, as it is evaluated from the direct numerical simulations for $\RA = 10^6$ and $\Gamma = 2$ (\textit{a}), $\Gamma = 2.5$ (\textit{b}), $\Gamma = 3$ (\textit{c}) and $\Gamma = 5$ (\textit{d}).
$\NU$ is calculated over the top and bottom plates, over the horizontal cross-section in the middle plane at $z=0.5H$ and over the entire volume.}
\label{fig_phaseavgheat_walls}
\end{figure}

In addition to figure \ref{fig_phaseavgheat} that shows the volume-averaged Nusselt number $\NU_{vol}$ during one oscillation period, we present in figure~\ref{fig_phaseavgheat_walls} the values of $\NU$, which are computed over the surfaces: $\NU_{bot}$ at the bottom plate, $\NU_{top}$ at the top plate and $\NU_{mid}$ over the horizontal cross-section in the middle plane at $z=0.5H$.
Compared to $\NU_{vol}$, for all studied values of $\Gamma$, there is a shift between $\NU$ evaluated at the plates ($\NU_{bot}$, $\NU_{top}$) and the volume-averaged Nusselt number $\NU_{vol}$.
The maxima and minima of $\NU$, calculated at the horizontal walls, occur always later than they appear in the $\NU_{vol}$ evolution.
$\NU_{bot}$ and $\NU_{top}$ are synchronised with each other. 
$\NU_{mid}$ seems to be less smooth and gives a larger difference between the maximum and minimum values compared to $\NU_{bot}$ and $\NU_{top}$.

\begin{figure}
%figure 14
\centerline{\includegraphics[scale = 1]{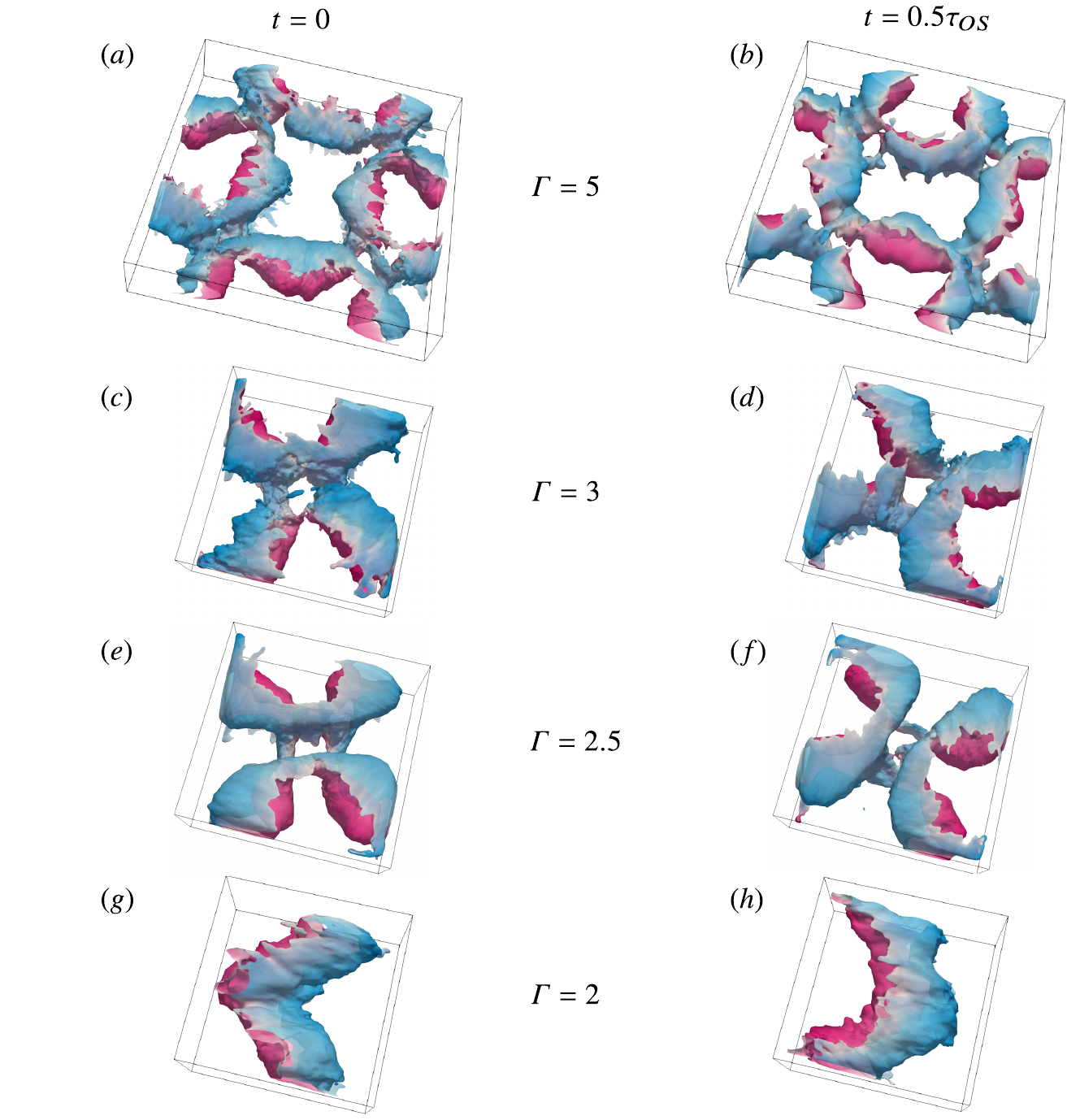}}
\caption{Isosurfaces of the magnitude of the full heat transport vector ${\vec{\Omega} \equiv (\vec{u}T - \kappa \vec{\nabla} T) /} (\kappa \Delta /H)$, for $\Gamma = 5$ (\textit{a,~b}) and $\Gamma = 3$ (\textit{c,~d}), $\Gamma = 2.5$ (\textit{e,~f}), $\Gamma = 2$ (\textit{g,~h}), as obtained in the simulations for $\RA = 10^6$. The surfaces are coloured by the temperature: blue (red) colour corresponds to the temperature below (above) the arithmetic mean of the top and bottom temperatures.}
\label{fig_heatflux_iso}
\end{figure}

Figure~\ref{fig_heatflux_iso} shows phase-averaged isosurfaces of the full heat transport vector $\vec{\Omega}$ as obtained in the direct numerical simulations for cuboid domains at the beginning ($t=0$) and at the middle ($t=0.5\tau_{OS}$) of the oscillation period.
The isosurfaces of the full heat transport vector ${\vec{\Omega}}$ follow the JRV flow structure at all considered $\Gamma$ values (cf. figure~\ref{streamline_vortex}).

\begin{figure}
%figure 15
\centerline{\includegraphics[scale = 1]{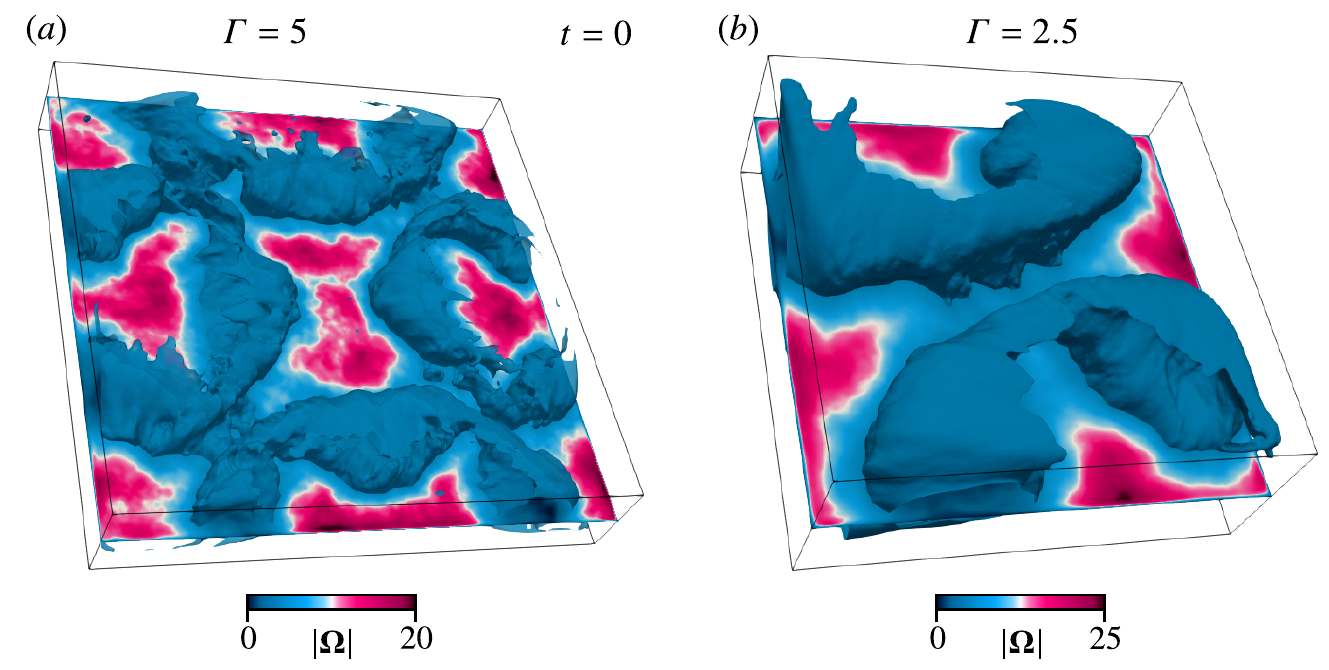}}
\caption{
Isosurfaces of the magnitude of the full heat transport vector $\vec{\Omega}$ are shown together with the  $|\vec{\Omega}|$ distribution at $z = 0.5 H$ for $\Gamma = 5$ (\textit{a}) and $\Gamma = 2.5$ (\textit{b}) for $\RA = 10^6$.
The JRV-like vortex structures are associated with the minimal heat flux. }
\label{fig_heatflux_iso_surf}
\end{figure}

\begin{figure}
%figure 16
\centerline{\includegraphics[scale = 1]{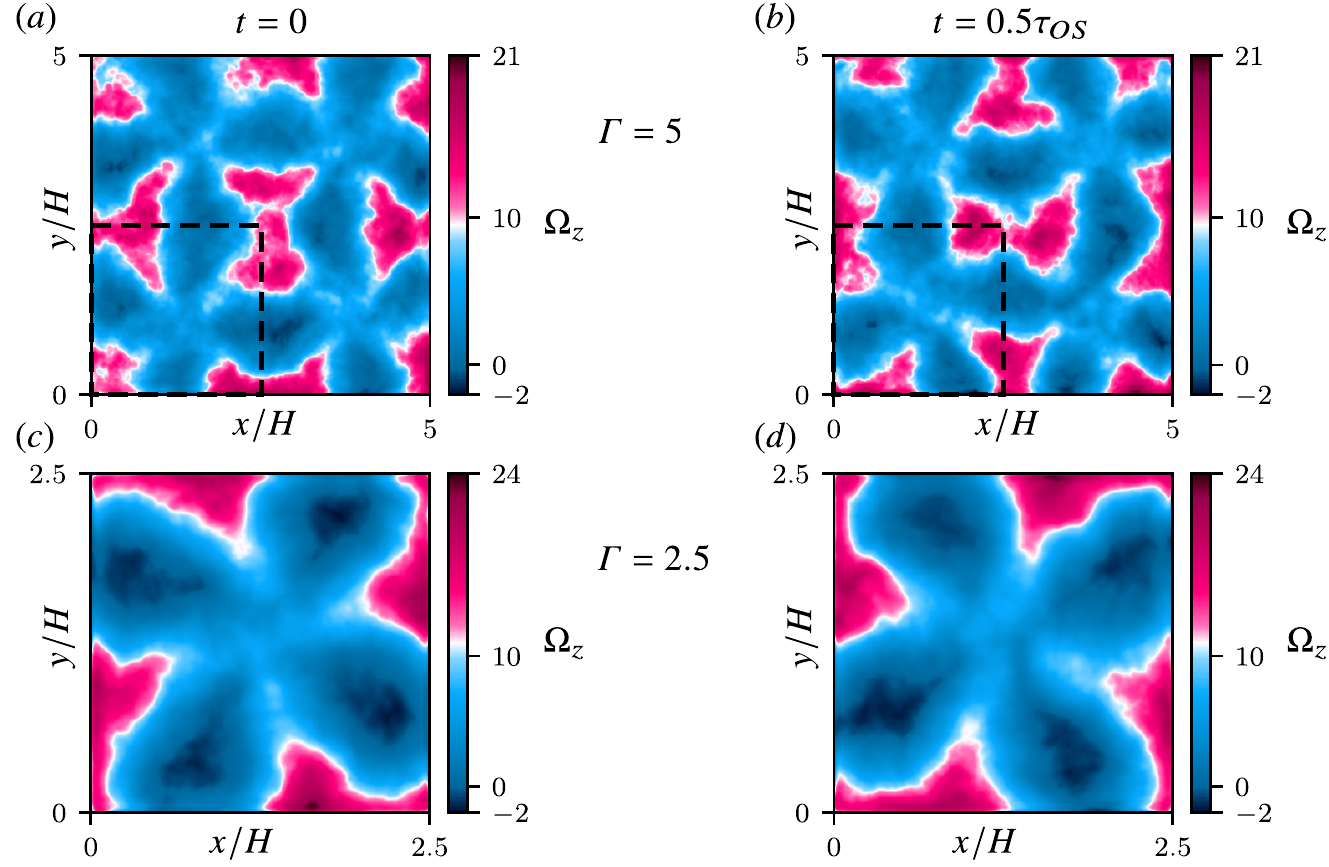}}
\caption{Phase-averaged vertical component of the local heat flux $\Omega_z$  at $z = 0.5 H$ for $\Gamma = 5$ (\textit{a,~b}) and $\Gamma = 2.5$ (\textit{c,~d}) for $\RA = 10^6$. 
The black squares indicate the areas that correspond to the areas of the container with $\Gamma = 2.5$ (see supplementary movies).
}
\label{fig_heatflux_surf}
\end{figure}

Figure~\ref{fig_heatflux_iso_surf} demonstrates ${\vec{\Omega}}$-isosurfaces together with the distribution of the magnitude, $|\vec{\Omega}|$, in the horizontal cross-section at $z = 0.5 H$ for $\Gamma = 5$ and $\Gamma = 2.5$.
The heat flow is mainly realized in the gaps between the isosurfaces that envelop the JRVs.
Thus, the JRVs are not efficient in transporting the heat and are located in the areas of minimum heat flux.

Figure~\ref{fig_heatflux_surf} shows the vertical  component of the local heat flux, $\Omega_z$,  at $z = 0.5 H$ for $\Gamma = 5$ and $\Gamma = 2.5$.
Analogously to figure~\ref{T_slices} with the temperature distributions, here one can see a similarity of the $\Omega_z$ distribution pattern in the case $\Gamma=2.5$ with the pattern in the 1/4 of the area in the case $\Gamma=5$ (see supplementary movies).
But the resemblance is not complete, probably because of the influence of the sidewalls.
How sidewalls affect the movements of vortices is a subject for future study.

\section{Discussion}

We have presented a combined numerical and experimental investigation of a liquid metal convection flow in different geometries. 
The Prandtl number in these investigations is $\Pran \approx 0.03$ and the Rayleigh numbers are between $2.9 \times 10^4 \leq \RA \leq 1.6 \times 10^6$. 
The investigations focus on the influence of the size of the flow domain (via its aspect ratio) on the dominant oscillation modes of the large-scale circulation. 
Results for four different cuboid domains with varying spatial length-to-height aspect ratios $\Gamma =5$, $\Gamma =3$, $\Gamma =2.5$ and $\Gamma =2$ were compared with the results of a cylindrical $\Gamma =2$ cell. 

The results show that the oscillations in all aspect ratios investigated are due to the presence of jump rope vortices. 
A jump rope vortex forms at the centre of the large-scale circulation, and moves analogous to a swirling jump rope. 
However, the direction of motion of the JRV is opposite to the direction of flow of the LSC. 
The JRV, which was first discovered in a cylindrical $\Gamma=2$ convection cell \citep{Vogt2018}, also forms in a square cuboid domain of aspect ratio, $\Gamma=2$, as demonstrated in this work. 
The appearance of the JRV is almost identical in both the cylindrical and cuboid domain of same aspect ratio. 
If a cylinder is numerically cut out from the rectangular cell, the similarity becomes more pronounced, also with respect to the JRV-induced sidewall temperature distribution. 

In domains with larger spatial length, the appearance of the JRV vortices changes. 
For domains with aspect ratios of $\Gamma=2.5$ and $\Gamma=3$, the vortices form an orthogonal cross that periodically rotates alternately clockwise and counterclockwise. 
In a $\Gamma=5$ cell, a lattice of four JRVs interlace each other, which oscillate in a synchronised manner. 
Therefore, a key finding of this work is that the JRV is an extremely robust flow feature that adapts and reorganises depending on different aspect ratio of a domain, with ability to form intricate lattice of repetitive flow structures in large aspect-ratio containers. 
Moreover, our findings further reinforce that the shape of the domain does matter: we encounter the presence of a JRV in a square cuboid of $\Gamma=3$, whereas, \citep{Cheng2022} did not find any evidence of a JRV in a cylinder of the same aspect ratio.
The frequency of the oscillations show a consistent scaling for the different aspect ratios with a good agreement between numerics and experiment. Slight deviations between the different aspect ratios are likely due to the non-uniform path length of the LSC for the different aspect ratios.

The heat transport scaling relations show only minor (if any) deviations between different flow pattern regimes. The data from the regime close to the onset, convection roll dominated regime and the turbulent JRV regime collapse on a master curve. However, the oscillations of the JRV are clearly visible in the time evolution of the Nusselt number. The frequency of the $\NU$ oscillations is thereby twice as high as that of the JRVs. The maxima of the Nusselt numbers occur when the horizontal velocity components reach a minima during the JRV cycle \citep[see~][]{Akashi2022}.

Questions, which are difficult to answer based on both experimental and numerical approaches, are whether the JRV structures have an upper $\RA$ number limit and whether they are subsequently displaced by other structures as soon as $\RA$ exceeds a certain critical value.
In the previous experiments, JRVs detected were stable over two orders of magnitude in $\RA$ \citep[see~][]{Vogt2018}. Since the flows in these measurements and simulations are already in a turbulent state, one might expect that the JRV-like oscillatory structures can occur for even larger $\RA$. 
It is worth noting that the JRVs occur not only for low Prandtl numbers like that we studied here, but have also been detected in a $\Gamma=2$ cylinder for water, which has approximately two orders of magnitude higher $\PR$ \citep{Horn2022}.

Our study poses few more questions for future studies that could potentially be investigated, such as: how do the JRVs behave in even larger containers with even higher spatial length domains, and what role do they play in formation of convective turbulent superstructures? 
The present study suggests that in the case of large $\Gamma$, the global structure of the oscillatory mode can be thought of as a lattice of interlaced JRV-like building blocks found for the aspect ratio $\Gamma\approx2.5$, repeated spatially. 
However, such investigations come with their own challenges. Numerical cost increases with the square of the domain aspect ratio, whereas the stabilizing influence of the sidewalls decreases with increasing aspect ratios, giving the flow more degrees of freedom, which results in JRVs that are less stable. This makes intractable the detection of the JRVs by known experimental techniques or by numerical techniques such as conditional averaging. Current ongoing research efforts at the HZDR aim to tackle this problem head-on by experimentally investigating the dynamics of oscillatory liquid metal thermal convection in a square cuboid with a large aspect ratio of $\Gamma=25$, which is under construction at the time of writing this paper.

\backsection[Supplementary data]{Supplementary material and movies are available at ...}

\backsection[Acknowledgements]{The authors thank Felix Schindler for assisting in the calibration of the set-up, and Susanne Horn for fruitful discussions.}

\backsection[Funding]{This work is supported by the Deutsche Forschungsgemeinschaft (DFG) under grants SH~405/16 and 
the Priority Programme SPP~1881 ``Turbulent Superstructures'' of the DFG under grants SH~405/7 and VO~2331/3.}

\backsection[Declaration of interests]{The authors report no conflict of interest.}

\backsection[Data availability statement]{The data that support the findings of this study are available under request.}

\backsection[Author ORCIDs]

Andrei Teimurazov https://orcid.org/0000-0002-2832-0335; 

Sanjay Singh https://orcid.org/0000-0002-5305-7524; 

Sylvie Su https://orcid.org/0000-0002-1794-1355; 

Sven Eckert https://orcid.org/0000-0003-1639-5417; 

Olga Shishkina https://orcid.org/0000-0002-6773-6464; 

Tobias Vogt https://orcid.org/0000-0002-0022-5758.

\backsection[Author contributions]{
A.~T. and S.~S. contributed equally to this work and should be considered joint first authors.
S.~S., S.~Su and A.~T. analysed the data. 
The numerical (experimental) part of the work was done by G\"ottingen (Dresden) group. 
Principle investigators of the project are O.~S. and T.~V. 
All authors contributed to the writing of the paper.
}

\bibliographystyle{jfm}
\bibliography{References}

\begin{thebibliography}{82}
\expandafter\ifx\csname natexlab\endcsname\relax\def\natexlab#1{#1}\fi
\def\au#1{#1} \def\ed#1{#1} \def\yr#1{#1}\def\at#1{#1}\def\jt#1{\textit{#1}}
  \def\bt#1{#1}\def\bvol#1{\textbf{#1}} \def\vol#1{#1} \def\pg#1{#1}
  \def\publ#1{#1}\def\arxiv#1{#1}\def\org#1{#1}\def\st#1{\textit{#1}}

\bibitem[Ahlers {\em et~al.\/}(2022)Ahlers, Bodenschatz, Hartmann, He, Lohse,
  Reiter, Stevens, Verzicco, Wedi, Weiss, Zhang, Zwirner \&
  Shishkina]{Ahlers2022}
{\sc \au{Ahlers, G.}, \au{Bodenschatz, E.}, \au{Hartmann, R.}, \au{He, X.},
  \au{Lohse, D.}, \au{Reiter, P.}, \au{Stevens, R.J.A.M.}, \au{Verzicco, R.},
  \au{Wedi, M.}, \au{Weiss, S.}, \au{Zhang, X.}, \au{Zwirner, L.} \&
  \au{Shishkina, O.}} \yr{2022}  \at{{Aspect ratio dependence of heat transfer
  in a cylindrical Rayleigh--B\'enard cell}}.  \jt{Phys. Rev. Lett.}
  \bvol{128},  \pg{084501}.

\bibitem[Ahlers {\em et~al.\/}(2009)Ahlers, Grossmann \& Lohse]{Ahlers2009}
{\sc \au{Ahlers, G.}, \au{Grossmann, S.} \& \au{Lohse, D.}} \yr{2009}
  \at{{Heat transfer and large scale dynamics in turbulent Rayleigh--B\'enard
  convection}}.  \jt{Rev. Mod. Phys.}  \bvol{81},  \pg{503--537}.

\bibitem[Akashi {\em et~al.\/}(2022)Akashi, Yanagisawa, Sakuraba, Schindler,
  Horn, Vogt \& Eckert]{Akashi2022}
{\sc \au{Akashi, M.}, \au{Yanagisawa, T.}, \au{Sakuraba, A.}, \au{Schindler,
  F.}, \au{Horn, S.}, \au{Vogt, T.} \& \au{Eckert, S.}} \yr{2022}  \at{{Jump
  rope vortex flow in liquid metal Rayleigh--B{\'e}nard convection in a cuboid
  container of aspect ratio}}.  \jt{J. Fluid Mech.}  \bvol{932},  \pg{A27}.

\bibitem[Akashi {\em et~al.\/}(2019)Akashi, Yanagisawa, Tasaka, Vogt, Murai \&
  Eckert]{Akashi2019}
{\sc \au{Akashi, M.}, \au{Yanagisawa, T.}, \au{Tasaka, Y.}, \au{Vogt, T.},
  \au{Murai, Y.} \& \au{Eckert, S.}} \yr{2019}  \at{{Transition from convection
  rolls to large-scale cellular structures in turbulent Rayleigh--B\'enard
  convection in a liquid metal layer}}.  \jt{Phys. Rev. Fluids}  \bvol{4},
  \pg{033501}.

\bibitem[Assaf {\em et~al.\/}(2011)Assaf, Angheluta \& Goldenfeld]{Assaf2011}
{\sc \au{Assaf, M.}, \au{Angheluta, L.} \& \au{Goldenfeld, N.}} \yr{2011}
  \at{Rare fluctuations and large-scale circulation cessations in turbulent
  convection}.  \jt{Phys. Rev. Lett.}  \bvol{107},  \pg{044502}.

\bibitem[B\'enard(1900)]{Benard1900}
{\sc \au{B\'enard, H.}} \yr{1900}  \at{Les tourbillons cellulairs dans une
  nappe liquide}.  \jt{Rev. G\'en. Sciences Pure Appl.}  \bvol{11},
  \pg{1261--1271, 1309--1328}.

\bibitem[Berghout {\em et~al.\/}(2021)Berghout, Baars \& Krug]{Berghout2021}
{\sc \au{Berghout, P.}, \au{Baars, W.} \& \au{Krug, D.}} \yr{2021}  \at{The
  large-scale footprint in small-scale rayleigh-b\'enard turbulence}.  \jt{J.
  Fluid Mech.}  \bvol{911},  \pg{A62}.

\bibitem[Bodenschatz {\em et~al.\/}(2000)Bodenschatz, Pesch \&
  Ahlers]{Bodenschatz2000}
{\sc \au{Bodenschatz, E.}, \au{Pesch, W.} \& \au{Ahlers, G.}} \yr{2000}
  \at{{Recent developments in Rayleigh--B\'enard convection}}.  \jt{Annu. Rev.
  Fluid Mech.}  \bvol{32},  \pg{709--778}.

\bibitem[Boffetta \& Ecke(2012)]{Boffetta2012}
{\sc \au{Boffetta, G.} \& \au{Ecke, R.~E.}} \yr{2012}  \at{Two-dimensional
  turbulence}.  \jt{Annu. Rev. Fluid Mech.}  \bvol{44},  \pg{427--451}.

\bibitem[Brown \& Ahlers(2006)]{Brown2006}
{\sc \au{Brown, E.} \& \au{Ahlers, G.}} \yr{2006}  \at{{Rotations and
  cessations of the large-scale circulation in turbulent Rayleigh--B\'enard
  convection}}.  \jt{J. Fluid Mech.}  \bvol{568},  \pg{351--386}.

\bibitem[Brown \& Ahlers(2007)]{Brown2007}
{\sc \au{Brown, E.} \& \au{Ahlers, G.}} \yr{2007}  \at{{Large-scale circulation
  model of turbulent Rayleigh--B\'enard convection}}.  \jt{Phys. Rev. Lett.}
  \bvol{98},  \pg{134501}.

\bibitem[Brown \& Ahlers(2009)]{Brown2009}
{\sc \au{Brown, E.} \& \au{Ahlers, G.}} \yr{2009}  \at{{The origin of
  oscillations of the large-scale circulation of turbulent Rayleigh--B\'enard
  convection}}.  \jt{J. Fluid Mech.}  \bvol{638},  \pg{383--400}.

\bibitem[Busse(1994)]{Busse1994}
{\sc \au{Busse, F.~H.}} \yr{1994}  \at{{Spoke pattern convection}}.  \jt{Acta
  Mechanica}  \bvol{4},  \pg{11--17}.

\bibitem[Cheng {\em et~al.\/}(2022)Cheng, Mohammad, Wang, Keogh, Forer \&
  Kelley]{Cheng2022}
{\sc \au{Cheng, J.~S.}, \au{Mohammad, I.}, \au{Wang, B.}, \au{Keogh, D.~F.},
  \au{Forer, J.~M.} \& \au{Kelley, D.~H.}} \yr{2022}  \at{Oscillations of the
  large-scale circulation in experimental liquid metal convection at aspect
  ratios 1.4-3}.  \jt{J. Fluid Mech.}  \bvol{949},  \pg{A42}.

\bibitem[Chill\`{a} \& Schumacher(2012)]{Chilla2012}
{\sc \au{Chill\`{a}, F.} \& \au{Schumacher, J.}} \yr{2012}  \at{{New
  perspectives in turbulent Rayleigh--B\'enard convection}}.  \jt{Eur. Phys. J.
  E}  \bvol{35},  \pg{58}.

\bibitem[Ching {\em et~al.\/}(2019)Ching, Leung, Zwirner \&
  Shishkina]{Ching2019}
{\sc \au{Ching, E. S.~C.}, \au{Leung, H.~S.}, \au{Zwirner, L.} \&
  \au{Shishkina, O.}} \yr{2019}  \at{{Velocity and thermal boundary layer
  equations for turbulent Rayleigh--B\'enard convection}}.  \jt{Phys. Rev.
  Res.}  \bvol{1},  \pg{033037}.

\bibitem[Cioni {\em et~al.\/}(1997)Cioni, Ciliberto \& Sommeria]{Cioni1997}
{\sc \au{Cioni, S.}, \au{Ciliberto, S.} \& \au{Sommeria, J.}} \yr{1997}
  \at{{Strongly turbulent Rayleigh--B\'enard convection in mercury: Comparison
  with results at moderate Prandtl number}}.  \jt{J. Fluid Mech.}  \bvol{335},
  \pg{111--140}.

\bibitem[Ecke \& Shishkina(2023)]{Ecke2023}
{\sc \au{Ecke, R.~E.} \& \au{Shishkina, O.}} \yr{2023}  \at{{Turbulent rotating
  Rayleigh--B\'enard convection}}.  \jt{Annu. Rev. Fluid Mech.}  \bvol{55},
  \pg{603--638}.

\bibitem[Eckert {\em et~al.\/}(2007)Eckert, Cramer \& Gerbeth]{Eckert2007}
{\sc \au{Eckert, S.}, \au{Cramer, A.} \& \au{Gerbeth, G.}} \yr{2007} {\em
  Velocity measurement techniques for liquid metal flows\/},  \st{Fluid
  Mechanics and Its Applications},  \vol{vol.~80}.  \publ{Springer}.

\bibitem[Emran \& Schumacher(2015)]{Emran2015}
{\sc \au{Emran, M.~S.} \& \au{Schumacher, J.}} \yr{2015}  \at{Large-scale mean
  patterns in turbulent convection}.  \jt{J. Fluid Mech.}  \bvol{776},
  \pg{96--108}.

\bibitem[Fitzjarrald(1976)]{Fitzjarrald1976}
{\sc \au{Fitzjarrald, D.~E.}} \yr{1976}  \at{{An experimental study of
  turbulent convection in air}}.  \jt{J. Fluid Mech.}  \bvol{73},
  \pg{693--719}.

\bibitem[Frick {\em et~al.\/}(2015)Frick, Khalilov, Kolesnichenko, Mamykin,
  Pakholkov, Pavlinov \& Rogozhkin]{Frick2015}
{\sc \au{Frick, P.}, \au{Khalilov, R.}, \au{Kolesnichenko, I.}, \au{Mamykin,
  A.}, \au{Pakholkov, V.}, \au{Pavlinov, A.} \& \au{Rogozhkin, S.~A.}}
  \yr{2015}  \at{{Turbulent convective heat transfer in a long cylinder with
  liquid sodium}}.  \jt{Europhys. Lett.}  \bvol{109},  \pg{14002}.

\bibitem[Funfschilling \& Ahlers(2004)]{Funfschilling2004}
{\sc \au{Funfschilling, D.} \& \au{Ahlers, G.}} \yr{2004}  \at{{Plume motion
  and large-scale dynamics in a cylindrical Rayleigh--B\'enard cell}}.
  \jt{Phy. Rev. Lett.}  \bvol{92},  \pg{194502}.

\bibitem[Funfschilling {\em et~al.\/}(2008)Funfschilling, Brown \&
  Ahlers]{Funfschilling2008}
{\sc \au{Funfschilling, D.}, \au{Brown, E.} \& \au{Ahlers, G.}} \yr{2008}
  \at{{Torsional oscillations of the large-scale circulation in turbulent
  Rayleigh--B\'enard convection}}.  \jt{J. Fluid Mech.}  \bvol{607},
  \pg{119--139}.

\bibitem[Green {\em et~al.\/}(2020)Green, Vlaykov, Mellado \&
  Wilczek]{Green2020}
{\sc \au{Green, G.}, \au{Vlaykov, D.~G.}, \au{Mellado, J.~P.} \& \au{Wilczek,
  M.}} \yr{2020}  \at{{Resolved energy budget of superstructures in
  Rayleigh--B\'enard convection}}.  \jt{J. Fluid Mech.}  \bvol{887},  \pg{A21}.

\bibitem[Grossmann \& Lohse(2000)]{Grossmann2000}
{\sc \au{Grossmann, S.} \& \au{Lohse, D.}} \yr{2000}  \at{{Scaling in thermal
  convection: A unifying theory}}.  \jt{J. Fluid Mech.}  \bvol{407},
  \pg{27--56}.

\bibitem[Grossmann \& Lohse(2001)]{Grossmann2001}
{\sc \au{Grossmann, S.} \& \au{Lohse, D.}} \yr{2001}  \at{{Thermal convection
  for large Prandtl numbers}}.  \jt{Phys. Rev. Lett.}  \bvol{86},
  \pg{3316--3319}.

\bibitem[Grossmann \& Lohse(2011)]{Grossmann2011}
{\sc \au{Grossmann, S.} \& \au{Lohse, D.}} \yr{2011}  \at{Multiple scaling in
  the ultimate regime of thermal convection}.  \jt{Phys. Fluids}  \bvol{23},
  \pg{045108}.

\bibitem[Hanasoge {\em et~al.\/}(2016)Hanasoge, Gizon \&
  Sreenivasan]{Hanasoge2016}
{\sc \au{Hanasoge, S.}, \au{Gizon, L.} \& \au{Sreenivasan, K.~R.}} \yr{2016}
  \at{{Seismic sounding of convection in the sun}}.  \jt{Ann. Rev. Fluid Mech.}
   \bvol{48},  \pg{191--217}.

\bibitem[von Hardenberg {\em et~al.\/}(2008)von Hardenberg, Parodi, Passoni,
  Provenzale \& Spiegel]{Hardenberg2008}
{\sc \au{von Hardenberg, J.}, \au{Parodi, A.}, \au{Passoni, G.},
  \au{Provenzale, A.} \& \au{Spiegel, E.~A.}} \yr{2008}  \at{{Large-scale
  patterns in Rayleigh-B\'enard convection}}.  \jt{Phys. Lett. A}  \bvol{372},
  \pg{2223}.

\bibitem[Hartlep {\em et~al.\/}(2003)Hartlep, Tilgner \& Busse]{Hartlep2003}
{\sc \au{Hartlep, T.}, \au{Tilgner, A.} \& \au{Busse, F.~H.}} \yr{2003}
  \at{{Large scale structures in Rayleigh--B\'enard convection at high Rayleigh
  numbers}}.  \jt{Phys. Rev. Lett.}  \bvol{91},  \pg{064501}.

\bibitem[Heinzel {\em et~al.\/}(2017)Heinzel, Hering, Konys, Marocco, Litfin,
  Mueller, Pacio, Schroer, Stieglitz, Stoppel, Weisenburger \&
  Wetzel]{Heinzel2017}
{\sc \au{Heinzel, A.}, \au{Hering, W.}, \au{Konys, J.}, \au{Marocco, L.},
  \au{Litfin, K.}, \au{Mueller, G.}, \au{Pacio, J.}, \au{Schroer, C.},
  \au{Stieglitz, R.}, \au{Stoppel, L.}, \au{Weisenburger, A.} \& \au{Wetzel,
  T.}} \yr{2017}  \at{Liquid metals as efficient high-temperature
  heat-transport fluids}.  \jt{Energy Technology}  \bvol{5},  \pg{1026--1036}.

\bibitem[Horn {\em et~al.\/}(2022)Horn, Schmid \& Aurnou]{Horn2022}
{\sc \au{Horn, S.}, \au{Schmid, P.~J.} \& \au{Aurnou, J.~M.}} \yr{2022}
  \at{{Unravelling the large-scale circulation modes in turbulent
  Rayleigh--B\'enard convection}}.  \jt{Europhys. Letters}  \bvol{136},
  \pg{14003}.

\bibitem[Kooij {\em et~al.\/}(2018)Kooij, Botchev, Frederix, Geurts, Horn,
  Lohse, van~der Poel, Shishkina, Stevens \& Verzicco]{Kooij2018}
{\sc \au{Kooij, G.~L.}, \au{Botchev, M.~A.}, \au{Frederix, E.~M.A.},
  \au{Geurts, B.~J.}, \au{Horn, S.}, \au{Lohse, D.}, \au{van~der Poel, E.~P.},
  \au{Shishkina, O.}, \au{Stevens, R. J. A.~M.} \& \au{Verzicco, R.}} \yr{2018}
   \at{{Comparison of computational codes for direct numerical simulations of
  turbulent Rayleigh--B\'enard convection}}.  \jt{Comp. Fluids}  \bvol{166},
  \pg{1--8}.

\bibitem[Krug {\em et~al.\/}(2020)Krug, Lohse \& Stevens]{Krug2020}
{\sc \au{Krug, D.}, \au{Lohse, D.} \& \au{Stevens, R. J. A.~M.}} \yr{2020}
  \at{{Coherence of temperature and velocity superstructures in turbulent
  Rayleigh--B\'enard flow}}.  \jt{J. Fluid Mech.}  \bvol{887},  \pg{A2}.

\bibitem[Li {\em et~al.\/}(2022)Li, Chen, Xu \& Xi]{Li2022}
{\sc \au{Li, Y.-Z.}, \au{Chen, X.}, \au{Xu, A.} \& \au{Xi, H.-D.}} \yr{2022}
  \at{Counter-flow orbiting of the vortex centre in turbulent thermal
  convection}.  \jt{J. Fluid Mech.}  \bvol{935},  \pg{A19}.

\bibitem[{Lord Rayleigh}(1916)]{Rayleigh1916}
{\sc \au{{Lord Rayleigh}}} \yr{1916}  \at{On convection currents in a
  horizontal layer of fluid, when the higher temperature is on the under side}.
   \jt{Phil. Mag.}  \bvol{32},  \pg{529--546}.

\bibitem[Pandey {\em et~al.\/}(2022)Pandey, Krasnov, Sreenivasan \&
  Schumacher]{Pandey2022}
{\sc \au{Pandey, A.}, \au{Krasnov, D.}, \au{Sreenivasan, K.} \& \au{Schumacher,
  J.}} \yr{2022}  \at{{Convective mesoscale turbulence at very low Prandtl
  numbers}}.  \jt{J. Fluid Mech.}  \bvol{948},  \pg{A23}.

\bibitem[Pandey {\em et~al.\/}(2018)Pandey, Scheel \& Schumacher]{Pandey2018}
{\sc \au{Pandey, A.}, \au{Scheel, J.~D.} \& \au{Schumacher, J.}} \yr{2018}
  \at{{Turbulent superstructures in Rayleigh--B\'enard convection}}.  \jt{Nat.
  Commun.}  \bvol{9},  \pg{2118}.

\bibitem[Plevachuk {\em et~al.\/}(2014)Plevachuk, Sklyarchuk, Eckert, Gerbeth
  \& Novakovic]{Plevachuk2014}
{\sc \au{Plevachuk, Y.}, \au{Sklyarchuk, V.}, \au{Eckert, S.}, \au{Gerbeth, G.}
  \& \au{Novakovic, R.}} \yr{2014}  \at{Thermophysical properties of the liquid
  ga--in--sn eutectic alloy}.  \jt{J. Chemical \& Eng. Data}  \bvol{59}~(3),
  \pg{757--763}.

\bibitem[van~der Poel {\em et~al.\/}(2011)van~der Poel, Stevens \&
  Lohse]{Poel2011}
{\sc \au{van~der Poel, E.~P.}, \au{Stevens, R. J. A.~M.} \& \au{Lohse, D.}}
  \yr{2011}  \at{{Connecting flow structures and heat flux in turbulent
  Rayleigh--B\'enard convection}}.  \jt{Phys. Rev. E}  \bvol{84},
  \pg{045303(R)}.

\bibitem[van~der Poel {\em et~al.\/}(2012)van~der Poel, Stevens, Sugiyama \&
  Lohse]{Poel2012}
{\sc \au{van~der Poel, E.~P.}, \au{Stevens, R. J. A.~M.}, \au{Sugiyama, K.} \&
  \au{Lohse, D.}} \yr{2012}  \at{{Flow states in two-dimensional
  Rayleigh--B\'enard convection as a function of aspect ratio and Rayleigh
  number}}.  \jt{Phys. Fluids}  \bvol{24},  \pg{085104}.

\bibitem[Reiter {\em et~al.\/}(2021)Reiter, Shishkina, Lohse \&
  Krug]{Reiter2021a}
{\sc \au{Reiter, P.}, \au{Shishkina, O.}, \au{Lohse, D.} \& \au{Krug, D.}}
  \yr{2021}  \at{{Crossover of the relative heat transport contributions of
  plume ejecting and impacting zones in turbulent Rayleigh--B\'nard
  convection}}.  \jt{Europhys. Lett.}  \bvol{134},  \pg{34002}.

\bibitem[Reiter {\em et~al.\/}(2022)Reiter, Zhang \& Shishkina]{Reiter2022}
{\sc \au{Reiter, P.}, \au{Zhang, X.} \& \au{Shishkina, O.}} \yr{2022}
  \at{{Flow states and heat transport in Rayleigh--B\'enard convection with
  different sidewall boundary conditions}}.  \jt{J. Fluid Mech.}  \bvol{936},
  \pg{A32}.

\bibitem[Sakievich {\em et~al.\/}(2016)Sakievich, Peet \&
  Adrian]{Sakievich2016}
{\sc \au{Sakievich, P.~J.}, \au{Peet, Y.~T.} \& \au{Adrian, R.~J.}} \yr{2016}
  \at{{Large-scale thermal motions of turbulent Rayleigh--B\'enard convection
  in a wide aspect-ratio cylindrical domain}}.  \jt{Int. J. Heat Fluid Flow}
  \bvol{61},  \pg{183--196}.

\bibitem[Sakievich {\em et~al.\/}(2020)Sakievich, Peet \&
  Adrian]{Sakievich2020}
{\sc \au{Sakievich, P.~J.}, \au{Peet, Y.~T.} \& \au{Adrian, R.~J.}} \yr{2020}
  \at{{Temporal dynamics of large-scale structures for turbulent
  Rayleigh--B\'enard convection in a moderate aspect-ratio cylinder}}.  \jt{J.
  Fluid Mech.}  \bvol{901},  \pg{A31}.

\bibitem[Scheel {\em et~al.\/}(2013)Scheel, Emran \& Schumacher]{Scheel2013}
{\sc \au{Scheel, J.~D.}, \au{Emran, M.~S.} \& \au{Schumacher, J.}} \yr{2013}
  \at{{Resolving the fine-scale structure in turbulent Rayleigh--B\'enard
  convection}}.  \jt{New J. Phys.}  \bvol{15},  \pg{113063}.

\bibitem[Scheel \& Schumacher(2016)]{Scheel2016}
{\sc \au{Scheel, J.~D.} \& \au{Schumacher, J.}} \yr{2016}  \at{{Global and
  local statistics in turbulent convection at low Prandtl numbers}}.  \jt{J.
  Fluid Mech.}  \bvol{802},  \pg{147--173}.

\bibitem[Schindler {\em et~al.\/}(2022)Schindler, Eckert, Z\"urner, Schumacher
  \& Vogt]{Schindler2022}
{\sc \au{Schindler, F.}, \au{Eckert, S.}, \au{Z\"urner, T.}, \au{Schumacher,
  J.} \& \au{Vogt, T.}} \yr{2022}  \at{Collapse of coherent large scale flow in
  strongly turbulent liquid metal convection}.  \jt{Phys. Rev. Lett.}
  \bvol{128},  \pg{164501}.

\bibitem[Schumacher {\em et~al.\/}(2015)Schumacher, G{\"o}tzfried \&
  Scheel]{Schumacher2015}
{\sc \au{Schumacher, J.}, \au{G{\"o}tzfried, P.} \& \au{Scheel, J.}} \yr{2015}
  \at{{Enhanced enstrophy generation for turbulent convection in
  low-Prandtl-number fluids}}.  \jt{Proc. National Acad. Sci.}
  \bvol{112}~(31),  \pg{9530--9535}.

\bibitem[Shishkina(2021)]{Shishkina2021}
{\sc \au{Shishkina, O.}} \yr{2021}  \at{{Rayleigh--B\'enard convection: The
  container shape matters}}.  \jt{Phys. Rev. Fluids}  \bvol{6},  \pg{090502}.

\bibitem[Shishkina \& Horn(2016)]{Shishkina2016b}
{\sc \au{Shishkina, O.} \& \au{Horn, S.}} \yr{2016}  \at{{Thermal convection in
  inclined cylindrical containers}}.  \jt{J. Fluid Mech.}  \bvol{790},
  \pg{R3}.

\bibitem[Shishkina {\em et~al.\/}(2015)Shishkina, Horn, Wagner \&
  Ching]{Shishkina2015}
{\sc \au{Shishkina, O.}, \au{Horn, S.}, \au{Wagner, S.} \& \au{Ching, E.
  S.~C.}} \yr{2015}  \at{{Thermal boundary layer equation for turbulent
  Rayleigh--B\'enard convection}}.  \jt{Phys. Rev. Lett.}  \bvol{114},
  \pg{114302}.

\bibitem[Shishkina {\em et~al.\/}(2010)Shishkina, Stevens, Grossmann \&
  Lohse]{Shishkina2010}
{\sc \au{Shishkina, O.}, \au{Stevens, R. J. A.~M.}, \au{Grossmann, S.} \&
  \au{Lohse, D.}} \yr{2010}  \at{Boundary layer structure in turbulent thermal
  convection and its consequences for the required numerical resolution}.
  \jt{New J. Phys.}  \bvol{12},  \pg{075022}.

\bibitem[Shishkina {\em et~al.\/}(2014)Shishkina, Wagner \&
  Horn]{Shishkina2014}
{\sc \au{Shishkina, O.}, \au{Wagner, S.} \& \au{Horn, S.}} \yr{2014}
  \at{Influence of the angle between the wind and the isothermal surfaces on
  the boundary layer structures in turbulent thermal convection}.  \jt{Phys.
  Rev. E}  \bvol{89},  \pg{033014}.

\bibitem[Spiegel(1962)]{Spiegel1962}
{\sc \au{Spiegel, E.~A.}} \yr{1962}  \at{{Thermal turbulence at very small
  Prandtl number}}.  \jt{J. Geophys. Res.}  \bvol{67},  \pg{3063--3070}.

\bibitem[Stevens {\em et~al.\/}(2018)Stevens, Blass, Zhu, Verzicco \&
  Lohse]{Stevens2018}
{\sc \au{Stevens, R. J. A.~M.}, \au{Blass, A.}, \au{Zhu, X.}, \au{Verzicco, R.}
  \& \au{Lohse, D.}} \yr{2018}  \at{{Turbulent thermal superstructures in
  Rayleigh--B\'enard convection}}.  \jt{Phys. Rev. Fluids}  \bvol{3},
  \pg{041501(R)}.

\bibitem[Stevens {\em et~al.\/}(2011)Stevens, Clercx \& Lohse]{Stevens2011}
{\sc \au{Stevens, R. J. A.~M.}, \au{Clercx, H. J.~H.} \& \au{Lohse, D.}}
  \yr{2011}  \at{{Effect of plumes on measuring the large scale circulation in
  turbulent Rayleigh--B\'enard convection}}.  \jt{Phys. Fluids}  \bvol{23},
  \pg{095110}.

\bibitem[Sugiyama {\em et~al.\/}(2010)Sugiyama, Ni, Stevens, Chan, Zhou, Xi,
  Sun, Grossmann, Xia \& Lohse]{Sugiyama2010}
{\sc \au{Sugiyama, K.}, \au{Ni, R.}, \au{Stevens, R. J. A.~M.}, \au{Chan,
  T.~S.}, \au{Zhou, S.-Q.}, \au{Xi, H.-D.}, \au{Sun, C.}, \au{Grossmann, S.},
  \au{Xia, K.-Q.} \& \au{Lohse, D.}} \yr{2010}  \at{Flow reversals in thermally
  driven turbulence}.  \jt{Phys. Rev. Lett.}  \bvol{105},  \pg{034503}.

\bibitem[Sun {\em et~al.\/}(2005{\natexlab{{\em a\/}}})Sun, Xi \&
  Xia]{Sun2005b}
{\sc \au{Sun, C.}, \au{Xi, H.-D.} \& \au{Xia, K.-Q.}} \yr{2005{\natexlab{{\em
  a\/}}}}  \at{Azimuthal symmetry, flow dynamics, and heat transport in
  turbulent thermal convection in a cylinder with an aspect ratio of 0.5}.
  \jt{Phys. Rev. Lett.}  \bvol{95},  \pg{074502}.

\bibitem[Sun {\em et~al.\/}(2005{\natexlab{{\em b\/}}})Sun, Xia \&
  Tong]{Sun2005}
{\sc \au{Sun, C.}, \au{Xia, K.-Q.} \& \au{Tong, P.}} \yr{2005{\natexlab{{\em
  b\/}}}}  \at{Three-dimensional flow structures and dynamics of turbulent
  thermal convection in a cylindrical cell}.  \jt{Phys. Rev. E}  \bvol{72},
  \pg{026302}.

\bibitem[Tai {\em et~al.\/}(2021)Tai, Ching, Zwirner \& Shishkina]{Tai2021}
{\sc \au{Tai, N.~C.}, \au{Ching, E. S.~C.}, \au{Zwirner, L.} \& \au{Shishkina,
  O.}} \yr{2021}  \at{{Heat flux in turbulent Rayleigh--B\'enard convection:
  Predictions derived from a boundary layer theory}}.  \jt{Phys. Rev. Fluids}
  \bvol{6},  \pg{033501}.

\bibitem[Teimurazov \& Frick(2017)]{Teimurazov2017}
{\sc \au{Teimurazov, A.} \& \au{Frick, P.}} \yr{2017}  \at{Thermal convection
  of liquid metal in a long inclined cylinder}.  \jt{Phys. Rev. Fluids}
  \bvol{2},  \pg{113501}.

\bibitem[Teimurazov {\em et~al.\/}(2017)Teimurazov, Frick \&
  Stefani]{Teimurazov2017a}
{\sc \au{Teimurazov, A.}, \au{Frick, P.} \& \au{Stefani, F.}} \yr{2017}
  \at{Thermal convection of liquid metal in the titanium reduction reactor}.
  \jt{IOP Conference Series: Materials Science and Engineering}  \bvol{208},
  \pg{012044}.

\bibitem[Teimurazov {\em et~al.\/}(2021)Teimurazov, Reiter, Shishkina \&
  Frick]{Teimurazov2021}
{\sc \au{Teimurazov, A.}, \au{Reiter, P.}, \au{Shishkina, O.} \& \au{Frick,
  P.}} \yr{2021}  \at{Heat transport in a cell heated at the bottom and the
  side}.  \jt{Europhys. Lett.}  \bvol{134}~(3),  \pg{34001}.

\bibitem[Tsuji {\em et~al.\/}(2005)Tsuji, Mizuno, Mashiko \& Sano]{Tsuji2005}
{\sc \au{Tsuji, Y.}, \au{Mizuno, T.}, \au{Mashiko, T.} \& \au{Sano, M.}}
  \yr{2005}  \at{Mean wind in convective turbulence of mercury}.  \jt{Phys.
  Rev. Lett.}  \bvol{94},  \pg{034501}.

\bibitem[Vogt {\em et~al.\/}(2021)Vogt, Horn \& Aurnou]{Vogt2021}
{\sc \au{Vogt, T.}, \au{Horn, S.} \& \au{Aurnou, J.~M.}} \yr{2021}
  \at{Oscillatory thermal–inertial flows in liquid metal rotating
  convection}.  \jt{J. Fluid Mech.}  \bvol{911},  \pg{A5}.

\bibitem[Vogt {\em et~al.\/}(2018)Vogt, Horn, Grannan \& Aurnou]{Vogt2018}
{\sc \au{Vogt, T.}, \au{Horn, S.}, \au{Grannan, A.~M.} \& \au{Aurnou, J.~M.}}
  \yr{2018}  \at{Jump rope vortex in liquid metal convection}.  \jt{Proc.
  National Acad. Sci.}  \bvol{115},  \pg{12674--12679}.

\bibitem[Wagner {\em et~al.\/}(2012)Wagner, Shishkina \& Wagner]{Wagner2012}
{\sc \au{Wagner, S.}, \au{Shishkina, O.} \& \au{Wagner, C.}} \yr{2012}
  \at{{Boundary layers and wind in cylindrical Rayleigh--B\'enard cells}}.
  \jt{J. Fluid Mech.}  \bvol{697},  \pg{336--366}.

\bibitem[Wang {\em et~al.\/}(2020)Wang, Verzicco, Lohse \& Shishkina]{Wang2020}
{\sc \au{Wang, Q.}, \au{Verzicco, R.}, \au{Lohse, D.} \& \au{Shishkina, O.}}
  \yr{2020}  \at{{Multiple states in turbulent large-aspect ratio thermal
  convection: What determines the number of convection rolls?}}  \jt{Phys. Rev.
  Lett.}  \bvol{125},  \pg{074501}.

\bibitem[Weiss \& Ahlers(2011)]{Weiss2011a}
{\sc \au{Weiss, S.} \& \au{Ahlers, G.}} \yr{2011}  \at{{Turbulent
  Rayleigh--B\'enard convection in a cylindrical container with aspect ratio
  $\Gamma=0.50$ and Prandtl number $Pr=4.38$}}.  \jt{J. Fluid Mech.}
  \bvol{676},  \pg{5--40}.

\bibitem[Weiss \& Ahlers(2013)]{Weiss2013}
{\sc \au{Weiss, S.} \& \au{Ahlers, G.}} \yr{2013}  \at{Effect of tilting on
  turbulent convection: cylindrical samples with aspect ratio $\gamma$=0.50}.
  \jt{J. Fluid Mech.}  \bvol{715},  \pg{314--334}.

\bibitem[Xi {\em et~al.\/}(2004)Xi, Lam \& Xia]{Xi2004}
{\sc \au{Xi, H.-D.}, \au{Lam, S.} \& \au{Xia, K.-Q.}} \yr{2004}  \at{{From
  laminar plumes to organized flows: The onset of large-scale circulation in
  turbulent thermal convection}}.  \jt{J. Fluid. Mech.}  \bvol{503},
  \pg{47--56}.

\bibitem[Xi \& Xia(2007)]{Xi2007}
{\sc \au{Xi, H.-D.} \& \au{Xia, K.-Q.}} \yr{2007}  \at{Cessations and reversals
  of the large-scale circulation in turbulent thermal convection}.  \jt{Phys.
  Rev. E}  \bvol{76},  \pg{036301}.

\bibitem[Xi \& Xia(2008)]{Xi2008}
{\sc \au{Xi, H.-D.} \& \au{Xia, K.-Q.}} \yr{2008}  \at{Flow mode transitions in
  turbulent thermal convection}.  \jt{Phys. Fluids}  \bvol{20},  \pg{055104}.

\bibitem[Xi {\em et~al.\/}(2006)Xi, Zhou \& Xia]{Xi2006}
{\sc \au{Xi, H.-D.}, \au{Zhou, Q.} \& \au{Xia, K.-Q.}} \yr{2006}  \at{Azimuthal
  motion of the mean wind in turbulent thermal convection}.  \jt{Phys. Rev. E}
  \bvol{73},  \pg{056312}.

\bibitem[Zhou {\em et~al.\/}(2009)Zhou, Xi, Zhou, Sun \& Xia]{Zhou2009}
{\sc \au{Zhou, Q.}, \au{Xi, H.-D.}, \au{Zhou, S.-Q.}, \au{Sun, C.} \& \au{Xia,
  K.-Q.}} \yr{2009}  \at{{Oscillations of the large-scale circulation in
  turbulent Rayleigh--B\'enard convection: The sloshing mode and its
  relationship with the torsional mode}}.  \jt{J. Fluid Mech.}  \bvol{630},
  \pg{367--390}.

\bibitem[Z\"urner {\em et~al.\/}(2019)Z\"urner, Schindler, Vogt, Eckert \&
  Schumacher]{Zuerner2019}
{\sc \au{Z\"urner, T.}, \au{Schindler, F.}, \au{Vogt, T.}, \au{Eckert, S.} \&
  \au{Schumacher, J.}} \yr{2019}  \at{Combined measurement of velocity and
  temperature in liquid metal convection}.  \jt{J. Fluid Mech.}  \bvol{876},
  \pg{1108--1128}.

\bibitem[Zwirner {\em et~al.\/}(2022)Zwirner, Emran, Schindler, Singh, Eckert,
  Vogt \& Shishkina]{Zwirner2022}
{\sc \au{Zwirner, L.}, \au{Emran, M.}, \au{Schindler, F.}, \au{Singh, S.},
  \au{Eckert, S.}, \au{Vogt, T.} \& \au{Shishkina, O.}} \yr{2022}
  \at{{Dynamics and length scales in vertical convection of liquid metals}}.
  \jt{J. Fluid Mech.}  \bvol{932},  \pg{A9}.

\bibitem[Zwirner {\em et~al.\/}(2020{\natexlab{{\em a\/}}})Zwirner, Khalilov,
  Kolesnichenko, Mamykin, Mandrykin, Pavlinov, Shestakov, Teimurazov, Frick \&
  Shishkina]{Zwirner2020a}
{\sc \au{Zwirner, L.}, \au{Khalilov, R.}, \au{Kolesnichenko, I.}, \au{Mamykin,
  A.}, \au{Mandrykin, S.}, \au{Pavlinov, A.}, \au{Shestakov, A.},
  \au{Teimurazov, A.}, \au{Frick, P.} \& \au{Shishkina, O.}}
  \yr{2020{\natexlab{{\em a\/}}}}  \at{{The influence of the cell inclination
  on the heat transport and large-scale circulation in liquid metal
  convection}}.  \jt{J. Fluid Mech.}  \bvol{884},  \pg{A18}.

\bibitem[Zwirner \& Shishkina(2018)]{Zwirner2018}
{\sc \au{Zwirner, L.} \& \au{Shishkina, O.}} \yr{2018}  \at{{Confined inclined
  thermal convection in low-Prandtl-number fluids}}.  \jt{J. Fluid Mech.}
  \bvol{850},  \pg{984--1008}.

\bibitem[Zwirner {\em et~al.\/}(2020{\natexlab{{\em b\/}}})Zwirner, Tilgner \&
  Shishkina]{Zwirner2020}
{\sc \au{Zwirner, L.}, \au{Tilgner, A.} \& \au{Shishkina, O.}}
  \yr{2020{\natexlab{{\em b\/}}}}  \at{{Elliptical instability and
  multiple-roll flow modes of the large-scale circulation in confined turbulent
  Rayleigh--B\'enard convection}}.  \jt{Phys. Rev. Lett.}  \bvol{125},
  \pg{054502}.

\end{thebibliography}

\end{document}